\documentclass[a4paper]{article}
\usepackage{epsfig}
\usepackage{graphicx}  

\date{\today}
\newcommand{\bmat}{\left(\begin{array}}
\newcommand{\emat}{\end{array}\right)}
\newcommand{\be}{\begin{equation}}
\newcommand{\ee}{\end{equation}}
\newcommand{\bea}{\begin{eqnarray}}
\newcommand{\eea}{\end{eqnarray}}

\def    \be            {\begin{equation}}
\def    \ee            {\end{equation}}
\def    \bea           {\begin{eqnarray}}
\def    \eea           {\end{eqnarray}}

\def\a{\alpha}
\def\b{\beta}

\def\d{\delta}
\def\n{\nu}

\def\lam{\lambda}
\def\th{\theta}

\def\m{\mu}
\def\nn{\nonumber}
\def\d{\delta}
\def\D{\Delta}
\def\s{\sigma}
\def\r{\rho}
\def\t{\theta}

\def\b{\beta}

\def\d{\delta}

\def\n{\nu}

\def\lam{\lambda}
\def\th{\theta}

\begin{document}
\title{\Large\bf The $U(1)$ symmetry of the non-tribimaximal pattern in the degenerate mass spectrum case of the neutrino mass matrix}
\author
{ \it \bf  E. I. Lashin$^{1,2,3}$\thanks{elashin@ictp.it} ,  N.
Chamoun$^{4,5}$\thanks{nchamoun@th.physik.uni-bonn.de}, and  S. Nasri$^{6}$ \thanks{snasri@uaeu.ac.ae} ,
\\ \small$^1$ Ain Shams University, Faculty of Science, Cairo 11566,
Egypt.\\
\small$^2$ Centre for Theoretical Physics, Zewail City of Science and Technology,\\
\small Sheikh Zayed, 6 October City, 12588, Giza, Egypt.\\
\small$^3$ The Abdus Salam ICTP, P.O. Box 586, 34100 Trieste, Italy.  \\
\small$^4$  Physics Department, HIAST, P.O.Box 31983, Damascus,
Syria. \\
 \small$^5$  Physikalisches Institut der Universit$\ddot{a}$t Bonn, Nu${\ss}$alle 12, D-53115 Bonn, Germany.
 \\
\small$^6$  Department of Physics, UAE University, P.O.Box 17551,
Al-Ain, United Arab Emirates.  }

\maketitle
\begin{center}
\small{\bf Abstract}\\[3mm]
\end{center}
On account of the new neutrino oscillation data signalling a non-zero value for the
 smallest mixing angle ($\theta_z$), we present an explicit realization of the underlying $U(1)$ symmetry characterizing the
maximal  atmospheric mixing angle ($\theta_y=\frac{\pi}{4}$) pattern with two degenerate masses but now with generic values of $\theta_z$.
We study the effects of the form invariance with respect to $U(1)$, and/or $Z_3$, $Z_2$ subgroups, on the Yukawa couplings and the mass terms.
Later on, we specify $\theta_z$ to its experimental best fit value ($ \sim 8^o$), and impose the symmetry in an entire model which includes
charged leptons, and many Higgs doublets or standard model singlet heavy scalars, to show that it can make room for the charged lepton mass hierarchies.
In addition, we show for the non-tribimaximal value of $\theta_z \neq 0$ within type-I seesaw mechanism enhanced with flavor symmetry that neutrino mass hierarchies can be generated. Furthermore, lepton/baryogenesis can be interpreted via type-II seesaw mechanism within a setup meeting the flavor $U(1)$-symmetry.
\\
{\bf Keywords}: Neutrino Physics; Flavor Symmetry; Matter-anti-matter.
\\
{\bf PACS numbers}: 14.60.Pq; 11.30.Hv; 98.80.Cq
\begin{minipage}[h]{14.0cm}
\end{minipage}
\vskip 0.3cm \hrule \vskip 0.5cm

\section{Introduction}
The lepton sector has a quite different pattern from that of the quarks mainly in two respects. First, the mixing among gauge eigenstates
in leptons to form mass eigenstates is more pronounced than in the quark sector, although the mass spectrum of the charged
leptons exhibits a similar hierarchy to the quarks' one \cite{nakamura10}. Second, the neutrino masses are quite small when compared with
all other mass scales, and this invokes some ways to understand this smallness, most popular of which is the seesaw mechanism within
grand unification \cite{seesaw}. As to the neutrino mass hierarchies, they are not yet determined experimentally, and many models based on
flavor symmetry considerations were constructed in order to account for the experimental data on neutrino masses and mixing (\cite{morisi} and references therein).


The `symmetric' neutrino mass matrix $M_\nu$
is diagonalized by a single unitary mixing matrix $U^\nu_L$ as follows
\bea
{M}_\nu &=& U^\nu_L\; M^{\mbox{diag}}_\n\; (U^\nu_L)^T,\;\;\;M^{\mbox{diag}}_\n = \mbox{diag}(m_1,m_2,m_3).
\label{neutrino mass}
\eea
 There are many possible parametrizations of the neutrino mixing matrix $U^\nu_L$, and we opt for the one in which the Dirac phase $\d$ does not appear in the effective mass term of the neutrinoless double decay \cite{Xing}. In this adopted parametrization, the mixing matrix $U^\nu_L$ is parameterized by three rotation angles ($\t_x,\t_y,\t_z$) and, in addition to $\d$,
two Majorana phases ($\r,\s$) as follows
\bea
U_L^\nu &=& R_{23}(\theta_y) \cdot R_z(\delta) \cdot R_{12}(\theta_x) \cdot P, \label{U decomposition} \nonumber
 \\
P = \mbox{diag}(e^{i\rho},e^{i\sigma},1), && R_z(\delta) =
\left ( \matrix{ c_z & 0  & s_z \cr
0 & e^{-i\,\d} & 0 \cr
-s_z & 0 &  c_z\cr } \right ),
 \label{U decomposition}
\eea
with $s_x \equiv \sin\theta_x$, $c_y \equiv \cos\theta_y$,  $t_z \equiv \tan \t_z$ (for later use), and so on. As to $R_{12}$ and $R_{23}$ they are rotations around the $z$ and $x$ axes respectively, while $m_i$'s are the masses of the neutrino mass states,
 leading to a mixing matrix of the form:
\bea U^\nu_L &=&  \left ( \matrix{ c_x c_z & s_x c_z & s_z \cr - c_x s_y
s_z - s_x c_y e^{-i\delta} & - s_x s_y s_z + c_x c_y e^{-i\delta}
& s_y c_z \cr - c_x c_y s_z + s_x s_y e^{-i\delta} & - s_x c_y s_z
- c_x s_y e^{-i\delta} & c_y c_z \cr } \right )\cdot \mbox{diag}(e^{i\rho},e^{i\sigma},1).
\label{mixmat}
\eea
There is a simple relation, discussed in \cite{oztex,ntb},  between this adopted parametrization and  the standard one used, say, in the recent
data analysis of \cite{fogli}.

In a similar way to the uncharged neutrinos case, the generally non-symmetric charged lepton mass matrix linking the left handed (LH) leptons to their
right handed (RH) counterparts can be diagonalized by a bi-unitary transformation:
\begin{equation}
{M}_l = U^l_L\;\cdot \mbox{diag}(m_e,m_\m,m_\tau)\;\cdot (U^l_R)^\dagger.
\end{equation}
We work in the flavor basis where $U_L^l = {\bf 1}$ the identity matrix, whence the flavor mixing matrix
$V \equiv (U^l_L)^\dagger\, U^\nu_L = U^\nu_L$, which can be constrained by observational data, comes wholly from
the neutrino sector in the flavor basis.

The authors of  \cite{HPS} noticed that the experimental data excluding the phases lead approximately to a specific pattern dubbed tri-bimaximal (TB):
\begin{eqnarray}
U^{\nu}_L \simeq V^{TB} &\equiv&
\pmatrix{\sqrt{2/3} & 1/\sqrt{3} & 0 \cr -1/\sqrt{6} & 1/\sqrt{3}
& 1/\sqrt{2} \cr 1/\sqrt{6} & -1/\sqrt{3} & +1/\sqrt{2}},
\end{eqnarray}
coming down to $\t_x=\arcsin{\left(\frac{1}{\sqrt{3}}\right)},\; \t_y={\pi \over 4}$ and $\t_z=0$.
 It has been shown in \cite{matripar}, that the TB pattern is equivalent to a certain form for the $M_\n$ in the flavor basis called `tripartite':
\bea
\label{equ-mass-tb-trip}
M^{\mbox{diag}}_\n =(V^{TB})^T \; M^{TB}_\n \;V^{TB} &\Leftrightarrow&  {M}^{TB}_{\nu} = { M}_A + { M}_B + { M}_C,
\eea
where
\begin{eqnarray} \label{tb-trip-form}
{M}_A = A \pmatrix {1 & 0 & 0 \cr 0 & 1 & 0 \cr 0 & 0 & 1}, ~~ {
M}_B = B \pmatrix {-1 & 0 & 0 \cr 0 & 0 & 1 \cr 0 & 1 & 0}, ~~ {
M}_C = C \pmatrix {1 & 1 & -1 \cr 1 & 1 & -1 \cr -1 & -1 & 1},
\end{eqnarray}
with neutrino eigen masses related to the tripartite coefficients via:
\begin{eqnarray}
\label{tri-par-eigenm}
m_1=A-B, & m_2 =A-B + 3C,&  m_3 = A+B, \nonumber\\
A=(m_1+m_3)/2, &  B=(m_3-m_1)/2, &  C=(m_2-m_1)/3.
\end{eqnarray}
Furthermore, a symmetry ($Z_3 \times Z_2$) for the `bipartite' form ($M_A+M_B$) corresponding to degenerate mass spectrum was given:
\bea
S^{TB}_3 = \pmatrix {-1/2 & -\sqrt{3/8} & \sqrt{3/8} \cr
\sqrt{3/8} & 1/4 & 3/4 \cr -\sqrt{3/8} & 3/4 & 1/4}:& & (S^{TB}_3)^3 =
{\bf 1},    \nn \\ S^{TB}_2 = \pmatrix {-1 & 0 & 0 \cr 0 & 0 & 1 \cr 0 & 1
& 0}: & & (S^{TB}_2)^2 = 1.
\label{masy}
\eea
The degenerate mass case corresponds to an ($O_2 \times Z^z_2$) symmetry, where the $O(2)$ group corresponds to the eigenspace
of the degenerate mass eigenvalue, whereas the $Z^z_2$ concerns the third mass value representing a reflection ($I_z=\mbox{diag}(1,1,-1)$) across its axis ($z$), whence
the superscript $z$. In \cite{u1}, we presented a specific realization of the `continuous' $U(1)$-symmetry characterizing the degenerate mass `bipartite' form ($M_A+M_B$) and studied its phenomenological consequences, be it in the corresponding current and conserved charges or in the possibility to
implement it in setups allowing for lepton mass hierarchies and baryogenesis.

Our method to find symmetry realizations characterizing the neutrino mass matrix in the flavor space consists in searching all the unitary
matrices $S$ satisfying what was called in \cite{maform} the form invariance formula:
\begin{equation}
\label{form}
S^T \;{M_\n}\;S = {M_\n}.
\end{equation}
The term $M_C$ in the tripartite form breaks the $U(1)$-part of $O(2)=U(1) \times Z^y_2$ \footnote{The $Z_2$ factor in the decomposition $O(2)=U(1)\times Z_2$ corresponds to the group consisting of the identity matrix and the reflection across one of the lines in the plane, say the y-axis, whence the notation $Z_2^y$.}, as we shall see, into $Z^\prime_2=\{I=\mbox{diag}(1,1,1),-I_z=\mbox{diag}(-1,-1,1)\}$, and we are left with a $Z^\prime_2 \times Z^y_2 \times Z^z_2$ discrete symmetry characterizing the most general non-degenerate symmetric neutrino mass matrix.  Moreover, since the expression ($S^T M_\n S$) is the same for a matrix ($S$) and its opposite ($-S$), we deduce that $Z^\prime_2$ is implied by $Z^z_2$ and the group $Z^y_2 \times Z^z_2$ characterizes uniquely the tripartite form. In addition, as $I_z \circ I_y = -I_x$ (where $I_i$ is the reflection across the $i$-axis) we find that the characterizing group $Z^y_2 \times Z^z_2$ is equivalent, as far as Eq.(\ref{form}) is concerned, to the discrete group ($Z_2^3$) corresponding to the three reflections ($I_x,I_y,I_z$) across
the axes in the diagonal basis. In \cite{z2}, we found specific realizations of this $Z_2^3$-symmetry in any basis defined by the mixing matrix $V$.

Although the TB pattern proved successful phenomenologically \cite{TB}, it was important to restudy these symmetries in light of the recent neutrino data departing from the TB pattern, in particular for the $\t_z$ angle whose vanishing TB value is no longer acceptable \cite{fogli,Double,DAYA,RENO, Valle}. We would designate the non-tribimaximal pattern with $\left[\t_y=\frac{\pi}{4}, \t_x = \arcsin{\left(\frac{1}{\sqrt{3}}\right)}
, \t_z \neq 0\right]$ the NTB pattern, assigning a subscript zero mark ($\mbox{NTB}_0$) for the case $\t_z=\arcsin{\left(\frac{1}{\sqrt{50}}\right)} \sim 8^o$ which
approximates well the best fit for $\t_z$. This rather large value $8^o$ renders implausible that the TB mixing pattern can be considered as a leading order approximation to be improved by perturbation towards the NTB$_0$ pattern. Thus the need arises to construct models that can accommodate large $\t_z$ from the outset.

In \cite{ntb}, we found the $Z_2^3$-symmetry realizations corresponding to the NTB pattern for generic $\t_z$, and studied the phenomenological consequences for the NTB$_0$ one. Since the degenerate mass spectrum case can be considered as a first step approximation perturbed by the $M_C$
part proportional to the neutrino mass splitting ($m_2-m_1$), it is of value to restudy this degenerate case but corresponding to the NTB pattern, which is just the objective of this paper. We note that the $\t_x$ angle is irrelevant in the degenerate case ($m_1=m_2$), since the corresponding
rotation matrix $R_{12}(\t_x)$ commutes with the degenerate mass matrix, so we designate any quantity corresponding to the NTB pattern
in the degenerate mass case by DNTB specified by ($\t_y=\frac{\pi}{4}, \t_z \neq 0, m_1=m_2$), whereas the special degenerate non-tribimaximal
pattern DNTB$_0$ is specified by $\left[\t_y=\frac{\pi}{4}, \t_z =\arcsin{\left(\frac{1}{\sqrt{50}}\right)}, m_1=m_2\right]$.

The plan of the paper is as follows. In subsection 2.1, of the section 2 devoted to the analysis of the $U(1)$-symmetry, we find an explicit
realization of the $U(1)$-symmetry, called henceforth the S-symmetry, leading to the DNTB pattern, and deduce the modified bipartite form characterizing it. We find also in this
 subsection the form of the Yukawa couplings dictated by the S-symmetry. In subsection 2.2, we infer the corresponding constraints
 on the forms of the mass matrix and the Yukawa couplings imposed by the $Z_3$ and $Z_2$ subgroups of $U(1)$, and justify the phenomenological
 equivalence between the continuous $U(1)$ S-symmetry and the $Z_3$ discrete symmetry. In subsection 2.3, we state all the preceding subsections' results but corresponding now to the `observed' DNTB$_0$ pattern, whose S-symmetry would be named $S_0$-symmetry. We use the latter results in section 3 within models containing charged leptons to show the
 possibility of generating the observed charged lepton mass hierarchies. The subsection 3.1 involves many Higgs doublets, while many standard model (SM) singlet heavy scalars are involved in subsection 3.2 keeping
 only one SM-Higgs doublet. We study also
 the current associated with the continuous S-symmetry in subsection 3.3. In section 4, we study the lepton S-symmetry in setups involving seesaw mechanisms. The subsection 4.1 involves type-I
 seesaw mechanism showing how to accommodate all kinds of neutrino mass hierarchies but with no lepton/baryogenesis. We need type-II seesaw mechanism enriched with flavor symmetry in order to account for baryogenesis in subsection 4.2. We end up with summary and conclusions in section 5.

 \section{\large \bf Analysis of the underlying symmetry of the DNTB pattern}
If we just restrict the atmospheric angle to its maximal value $\left(\t_y=\frac{\pi}{4}\right)$ leaving $\t_x$ and $\t_z$ general, we get the mixing matrix (ignoring phases):
\bea
V^{xz} &=& \pmatrix{ c_z c_x &  c_z s_x & s_z  \cr  -\frac{1}{\sqrt{2}} ( s_z c_x + s_x)
& -\frac{1}{\sqrt{2}} ( s_z s_x - c_x) & \frac{1}{\sqrt{2}} c_z \cr  -\frac{1}{\sqrt{2}} ( s_z c_x - s_x) & -\frac{1}{\sqrt{2}} ( s_z s_x + c_x)
& \frac{1}{\sqrt{2}} c_z}.
\label{Vxz}
\eea
Then the neutrino mass matrix in Eq.~(\ref{neutrino mass}) can be cast in a general tripartite form:
\be
{\large
\begin{array}{lll}
M_{\n} &=& \pmatrix{A-B+C &
\frac{\sqrt{2}c_z s_z}{1-3s_z^2}\,B + \frac{\left(s_z c_{2x} - \frac{1}{2} s_{2x} c_z^2 \right)c_z}{\sqrt{2} s_x\, c_x\, s_z \left(1-3s_z^2\right)}\,C  & \frac{\sqrt{2}\,c_z\, s_z}{1-3s_z^2}\,B + \frac{s_z c_{2x} +  \frac{1}{2} s_{2x} \left(1-5s_z^2\right)}{\sqrt{2}\, s_x\, c_x\, s_z \left(1-3s_z^2\right)}\,C\\
\cr -  & A+C &  \frac{c^2_z}{1-3s_z^2}\,B + \frac{c_{2z}\, c_{2x} - s_z\, s_x\, c_x\, c_z^2}{ s_x\, c_x\, s_z\, \left(1-3s_z^2\right)}\,C\\
\cr  - &  - & A-C},
\label{gen-tri-for}
\end{array}
}
\ee
where the missed entries from this point on are determined from the matrix being symmetric. The coefficients of this general tripartite form are given in terms of $\t_x$, $\t_z$ and the neutrino masses as follows.
\bea
\label{gen-tri-par-coe}
A&=& \frac{1}{2}\, c_z^2 \left(\frac{1}{2}-c_x^2\right) \left(m_1-m_2\right) + \frac{1}{2} \left(1-\frac{1}{2}\,c_z^2\right) \left(m_1+m_2\right) + \frac{1}{2}\, c_z^2 m_3,\nonumber\\
B&=& \frac{1}{2}\, \left[-3 c_z^2 \left(-\frac{1}{2}+c_x^2\right) + 2 s_z s_x c_x \right] \left(m_1-m_2\right) + \frac{1}{2}\, \left(1-\frac{3}{2} c_z^2\right)\,\left(m_1+m_2\right) + \left(\frac{3}{2}\, c_z^2 -1\right)\, m_3,\nonumber\\
C&=& s_x c_x s_z \left(m_1-m_2\right).
\eea

We restrict our study henceforth to the degenerate non-tribimaximal pattern DNTB specified by $\left(\t_y=\frac{\pi}{4}, m_1=m_2\right)$. We see
directly from Eq.(\ref{gen-tri-par-coe}) that the $\t_x$-dependence is dropped, as well as the perturbation part involving $C$, and
we are left with a modified symmetric generic bipartite form:
 \bea
\label{gen-bip-for}
M^{DNTB}_{\n} &=& \pmatrix{A-B & \frac{\sqrt{2} s_z c_z}{1-3s_z^2} B & \frac{\sqrt{2} s_z c_z}{1-3s_z^2} B
\cr - & A&  \frac{c_z^2}{1-3s_z^2} B\cr
- &  - & A},
\eea
where the neutrino eigen masses and the coefficients of this  generic bipartite form are related by:
\bea
\label{gen-deg-eigen-mass}
m_1 =m_2 =  A-\frac{c_z^2}{1-3s_z^2}\,B,& & A = \left(1-\frac{1}{2}\,c_z^2\right)\, m_1 + \frac{1}{2}\, c_z^2 m_3
\nonumber, \\
m_3 = A+\frac{1+s_z^2}{1-3s_z^2}\,B,& & B = \left(1-\frac{3}{2}\,c_z^2\right) m_1 + \left(\frac{3}{2}\, c_z^2 -1\right)\, m_3 .
\eea
The mass spectrum of this generic bipartite form of the neutrino mass matrix
can account for all types of neutrino mass hierarchies as follows (assuming small $\t_z$):
\begin{itemize}
\item{ Normal hierarchy}:
\begin{eqnarray}
 A \simeq B &\Rightarrow& m_1 = m_2 \ll m_3, \label{nor-hier}
\end{eqnarray}

\item{ Inverted hierarchy}:
\begin{eqnarray}
A \simeq - B&\Rightarrow& m_3 \ll m_1 = m_2, \label{inv-hier}
\end{eqnarray}

\item{ Degenerate case}:
\begin{eqnarray}
A \gg B &\Rightarrow& m_1 = m_2 \simeq m_3, \label{deg-hier}
\end{eqnarray}
\end{itemize}

Some `mass relations', characterizing the non-triplicity in the degenerate case, can be deduced here like:
\bea
\label{gen-deg-mass-rel}
M_\n^{\mbox{\textsc{dntb}}} (1,2)  &=& M_\n^{\mbox{\textsc{dntb}}} (1,3),\nonumber \\
M_\n^{\mbox{\textsc{dntb}}} (1,2)  &=& \sqrt{2}\, t_z\,  M_\n^{\mbox{\textsc{dntb}}} (2,3),\nonumber \\
M_\n^{\mbox{\textsc{dntb}}} (1,1) &=& M_\n^{\mbox{\textsc{dntb}}} (2,2) - \frac{1-3s_z^2}{\sqrt{2} s_z c_z} M_\n^{\mbox{\textsc{dntb}}} (1,2), \nonumber \\
 M_\n^{\mbox{\textsc{dntb}}} (1,1) &=& M_\n^{\mbox{\textsc{dntb}}} (2,2) - \frac{1-3s_z^2}{c^2_z} M_\n^{\mbox{\textsc{dntb}}} (2,3),
\eea
 which for vanishing $\t_z$ reduce, as expected, to  the simpler ones characterizing the triplicity in the degenerate case, namely,
\bea
M_\n^{\mbox{\textsc{dtb}}}(1,2) = M_\n^{\mbox{\textsc{dtb}}}(1,3) &=& 0,\nonumber \\
 M_\n^{\mbox{\textsc{dtb}}}(2,2) - M_\n^{\mbox{\textsc{dtb}}}(2,3) &=& M_\n^{\mbox{\textsc{dtb}}}(1,1).
\label{deg-tb-mass-rel}
\eea

Before carrying out the analysis, it is important at this stage to quantify the $C$-term (Eq. \ref{gen-tri-par-coe}) breaking the $O(2)$-symmetry, which is proportional to the
mass splitting ($m_2-m_1 \geq 0$), and estimate its relative size with respect to other terms conserving this symmetry, and which in the eigen-space
corresponding to $m_1$ and $m_2$ are proportional to ($m_1+m_2$). In this way,  a small value of the ratio
\be r=\frac{m_2-m_1}{m_2+m_1}\label{r}\ee would indicate
that the $O(2)$-symmetry is satisfied to a good approximation.

It is worthy
to note here that, in many theoretically well justified and experimentally acceptable patterns for the neutrino mass matrix, numerical outcomes
led to $\frac{m_2}{m_1}<1.05$ implying $r< 1-\frac{m_1}{m_2} < 5\%$ \cite{oztex}. In  a model-independent way, we write down in Table \ref{tab1}
the latest global-fit carried out in \cite{fogli} for the mixing angles and the solar ($\d m^2$) and atmospheric ($|\D m^2|$) mass-squared differences defined by
 \begin{equation}
\delta m^2 \; \equiv \; m_2^2-m_1^2 \; , \; \left|\Delta m^2\right|  \;\equiv \; \left|m_3^2-{1\over 2}\left(m_1^2+m_2^2\right)\right|\;\; ,
\label{sqm}
\end{equation}
and also for the parameter ($R_\n$) characterizing the hierarchy between these two quantities:
\begin{equation}
R_\nu \; \equiv \;  \frac{\delta m^2} {\left|\Delta
m^2\right|}.
\label{rnu}
\end{equation}

 \begin{table}[h]
 \begin{center}
{\small
 \begin{tabular}{c c c }
\hline
\hline
\mbox{Parameter} &$\mbox{Best fit}$  & $2\s$ \mbox{range} \\
\hline
 $\d m^2 (10^{-5}\mbox{eV}^2)$ &$7.58$ &$\left[7.16, 7.99\right]$  \\
 \hline
 $\left|\D m^2\right|(10^{-3}\mbox{eV}^2)$& $2.35$  &$\left[2.17, 2.57\right]$  \\
 \hline
$\th_x$ & $33.58^o$  &$\left[31.95^o, 36.09^o\right]$  \\
\hline
$\th_y$ &$40.40^o$  &$\left[36.87^o, 50.77^o\right]$ \\
\hline
$\th_z$ & $8.33^o$  &$\left[6.29^o, 11.68^o\right]$  \\
\hline
$R_{\nu}$ & $0.0323$ &$\left[0.0279 , 0.0368\right]$ \\
\hline
 \end{tabular}
 }
 \end{center}
  \caption{\small \label{tab1} The latest global-fit results of the three neutrino mixing angles $(\th_x, \th_y, \th_z)$ and the two neutrino mass-squared differences $\d m^2$ and $\D m^2$ as defined in Eq.~(\ref{sqm}). Here, it is assumed that $\cos{\d}= \pm 1$ and that new reactor fluxes have been used\cite{fogli}.}
 \end{table}

We can estimate the ratio ($r=\frac{\d m^2}{(m_1+m_2)^2}$) in all the three types of neutrino mass hierarchies as follows. In the normal hierarchy
($m_1\leq m_2 \ll m_3$), we
note that the experimental data excluding two vanishing neutrino masses forbid, in the case of degenerate mass spectrum, a zero value for $m_1$ corresponding to ($r=1$). However, even when approaching this `extreme' case of $m_1 \sim 0$, the part proportional to $m_3$ in the neutrino mass matrix (Eqs~ \ref{gen-tri-for},\ref{gen-tri-par-coe}), and which does not affect the $O(2)$-symmetry, would be preponderant compared to the $O(2)$-conserving term proportional to ($m_2+m_1$) and the  $O(2)$-breaking term proportional to ($m_2-m_1$), as we have here,
 \be
 (m_2+m_1)/ m_3 \sim (m_2-m_1)/ m_3 \sim m_2/m_3 \sim {\cal O}\left(\sqrt{\d m^2/\D m^2}\;\right) \sim {\cal O}\left(\sqrt{R_\nu}\;\right) \sim 18\%.
 \ee
For `large' non-vanishing value of $m_1$, the ratio $r$ can take quite small values, and as an estimate we evaluate $r$ when $m_1^2 \sim \d m^2$ leading to $m_2^2 \sim 2 \d m^2$, whence $r \sim 1/(1+\sqrt{2})^2 < 18\%$. The larger $m_1$, the smaller $r$, so for example when $(m_1+m_2)^2$ is of order {$\cal{O}$}($|\D m^2|$), then $r={\cal O}(R_\n)={\cal O}(3\%)$.
In the inverted hierarchy case ($m_3 \ll m_1 \leq m_2$), one can estimate ($m_3$) by a very tiny value, so we get ($m_1 \sim m_2 \sim \sqrt{|\D m^2|} $),
 and so we get $r \sim R_\n /4 \sim 0.8\%$. Finally, in the degeneracy spectrum case ($m_1 \sim m_2 \sim m_3 \sim m_0$), we should have ($m_0 \geq \sqrt{|\D m^2|}$),
 and so $r<0.8\%$. Thus we can say that both experimental and numerical results corroborate the degenerate mass case as a good starting approximation for the non-degenerate spectrum case.

\subsection{\large \bf $U(1)$-symmetry in the {\sc{dntb}} pattern}
In order to find the symmetries imposing the form of Eq.(\ref{gen-bip-for}), we see that any unitary matrix $U$ satisfying the form invariance
in the `diagonalized' basis:
\begin{equation}
\label{deg-form-diag}
U^{\mbox{\textsc{t}}} \cdot M^{\mbox{\textsc{d},diag}}_\n \cdot U = M^{\mbox{\textsc{d},diag}}_\n \equiv \mbox{diag}(m_1, m_1, m_3),
\end{equation}
corresponds to a unitary matrix
\bea
\label{SDNTB}
S^{\mbox{\textsc{dntb}}} &=& \left(V^{\mbox{xz}}\right)^* \cdot U \cdot \left(V^{\mbox{xz}}\right)^{\mbox{\textsc{t}}},
\eea
satisfying the form invariance (Eq.\ref{form}) in the
degenerate mass spectrum case:
\be
\label{deg-form-non-diag}
(S^{\mbox{\textsc{dntb}}})^{\mbox{\textsc{t}}} \cdot M^{\mbox{\textsc{dntb}}}_\n \cdot S^{\mbox{\textsc{dntb}}} = M^{\mbox{\textsc{dntb}}}_\n,
\ee
where
\be
M^{\mbox{\textsc{d},diag}}_\n= \left(V^{\mbox{xz}}\right)^{\mbox{\textsc{t}}}\; \cdot M^{\mbox{\textsc{dntb}}}_\n \cdot V^{\mbox{xz}} .
\ee
It is clear now that the unitary matrices $U$ satisfying Eq.(\ref{deg-form-diag}) represent a group $O(2) \times Z_2$, where
$Z_2=\{\mbox{I},\,\mbox{I}_z\}$, while the orthogonal group $O(2)$ is a direct product
of rotations ($SO(2) \cong U(1)$) in the degenerate eigenspace and another $Z_2$ representing a reflection in this space. If,
 for continuity purposes, we restrict ourselves to the
connected component of the unity, then we have:
\bea
\label{U}
U(\t)=R_{12}(\t) &\equiv& \pmatrix{c_\t & s_\t & 0  \cr -s_\t & c_\t
&  0\cr  0 &  0 & 1}.
\eea
We thus deduce the `continuous' S-symmetry in the flavor basis by applying Eq.(\ref{SDNTB}):
 \bea
 \label{S-DNTB-sym}
 S^{\mbox{\textsc{dntb}}}_\t&=& \pmatrix{c_\t\, c_z^2 + s_z^2 & -\frac{c_z}{\sqrt{2}}\,\left[-s_\t+s_z\left(c_\t-1\right)\right] &  -\frac{c_z}{\sqrt{2}}\left[s_\t+s_z\left(c_\t-1\right)\right] \cr -\frac{c_z}{\sqrt{2}}\left[s_\t+s_z\left(c_\t-1\right)\right] & c_\t -\frac{1}{2}c_z^2\, \left(c_\t-1\right)
&  s_\t s_z -\frac{1}{2}c_z^2 \left(c_\t-1\right)\cr  -\frac{c_z}{\sqrt{2}}\left[-s_\t+s_z\,\left(c_\t-1\right)\right] &  -s_\t s_z -\frac{1}{2} c_z^2\, \left(c_\t-1\right) & c_\t -\frac{1}{2}\,c_z^2\, \left(c_\t-1\right)}.
\eea
We note that the $\t_x$-dependence in Eq.(\ref{SDNTB}) disappears since the two rotations around the third axis $R_{12}(\t_x)$ and $R_{12}(\t)$ commute.
One can check now that this S-symmetry is equivalent to the DNTB modified generic bipartite form in that for all angles $\t_z$ we have the following:
\bea
\label{u1-gen-mass-equivalence}
\left\{ \left(M=M^{\mbox{\textsc{t}}}\right)\,\wedge \left[\forall \t, (S^{\mbox{\textsc{dntb}}}_\t)^{\mbox{\textsc{t}}} \cdot M \cdot S^{\mbox{\textsc{dntb}}}_\t=M \right] \right\}
& \Leftrightarrow & { \left[ \exists\, A,B,C: M = \pmatrix{A-B & \frac{\sqrt{2} s_z c_z}{1-3s_z^2} B & \frac{\sqrt{2} s_z c_z}{1-3s_z^2} B
\cr - & A &  \frac{c_z^2}{1-3s_z^2} B\cr
- &  - & A}
\right]},\nonumber\\
\eea
If we drop the symmetric matrix condition in Eq.(\ref{u1-gen-mass-equivalence}) then we get for all angles $\t_z$ the following equivalence (which will prove useful when studying the general form of the Yukawa couplings in the adopted Lagrangian with specific fields transformations under S-symmetry):
\be
\label{u1-gen-yuk-equivalence}
\begin{array}{lll}
\left[  \forall \t, \left(S^{\mbox{\textsc{dntb}}}_\t\right)^{\mbox{\textsc{t}}} \cdot f \cdot S^{\mbox{\textsc{dntb}}}_\t=f  \right]
 \Leftrightarrow &&\nonumber\\\\
  { \left[ \exists\, A,B,C: f = \pmatrix{A-B+\sqrt{2}\,t_z\, C & C
& \frac{\sqrt{2} s_{2z}}{1-3s_z^2}\, B - \frac{1+ s^2_z}{1-3s_z^2}\, C
\cr \frac{\sqrt{2} s_{2z}}{1-3s_z^2}\, B - \frac{1+ s^2_z}{1-3s_z^2}\, C & A
&  B\cr  C &    \frac{1+ s^2_z}{1-3s_z^2}\, B - \frac{\sqrt{2} s_{2z}}{1-3s_z^2}\, C& A}
\right]},&&\nonumber\\
\end{array}
\ee
Note that the equivalence Eq.(\ref{u1-gen-mass-equivalence}) can be deduced from that of Eq.(\ref{u1-gen-yuk-equivalence})
by the following substitution
\bea
B\rightarrow\frac{c_z^2}{1-3s_z^2}B, && C\rightarrow \frac{\sqrt{2} s_z c_z}{1-3s_z^2}B.
\eea

Also, it turns out useful to have the following equivalence corresponding to a `left-congruous' form invariance:
\bea
\label{u1-gen-yuk-lef-equivalence}
\left[  \forall \t, \left(S^{\mbox{\textsc{dntb}}}_\t\right)^{\mbox{\textsc{t}}} \cdot f =f  \right]
& \Leftrightarrow & { \left[ \exists\, A,B,C: f = \pmatrix{\sqrt{2}\,t_z \,A & \sqrt{2}\,t_z\, B
& \sqrt{2}\,t_z\, C
\cr A & B
&  C\cr  A &    B& C}
\right]},
\eea

One last note in this subsection is that we have neglected the Majorana phases in our discussion so far. However, diagonalizing the modified
generic bipartite form for the neutrino mass matrix in the DNTB pattern, we have:
\bea
\pmatrix{A-B & \frac{\sqrt{2} s_z c_z}{1-3s_z^2} B & \frac{\sqrt{2} s_z c_z}{1-3s_z^2} B
\cr - & A&  \frac{c_z^2}{1-3s_z^2} B\cr
- &  - & A} = &V^{x}_0 \cdot P \cdot \mbox{diag} \left( |A-\frac{c_z^2}{1-3s_z^2}B|,|A-\frac{c_z^2}{1-3s_z^2}B|,|A+\frac{1+s_z^2}{1-3s_z^2}B| \right) \cdot P^T \cdot (V^x_0)^T& \nonumber \\
\eea
where $P$, a diagonal phase matrix, is given by
 \bea
 P=\mbox{diag}\left(e^{i\a}, e^{i\a}, e^{i\b}\right), && 2\,\a= \arg\left(A-\frac{c_z^2}{1-3s_z^2}\,B\right),\, 2\,\b=
 \arg \left(A+\frac{1+s_z^2}{1-3s_z^2}\,B\right).
 \eea
We can absorb the $\b$ phase by an ``unphysical'' global phase shift of the neutrino fields ($\n_i \rightarrow e^{-i\b} \n_i$)
in the neutrino mass term $(M^{\mbox{\textsc{dntb}}_0}_{ij} \n_i \n_j)$, so when we compare with Eqs. (\ref{neutrino mass}-\ref{mixmat}) we find,
\be
\r=\s,\;\d=0,\;\t_y=\pi/4.
\ee
Conversely, starting from the following general expressions of the elements of the degenerate mass matrix (resulting from Eqs. \ref{neutrino mass}-\ref{mixmat} with $m_1=m_2$):
{\small
\bea \label{degenerate matrix elements}
&& M_{\n\,11}= m_1 c_z^2\, \left(c_x^2\,e^{2i\r} +s_x^2\,e^{2i\s}\right)
+ m_3 s_z^2, \nn\\
&&M_{\n\,12}=  m_1 \left[ - c_z s_z s_y \left(c_x^2 e^{2i\r} + s_x^2 e^{2i\s}\right) + c_z c_x s_x c_y \left(e^{i(2\s-\d)}- e^{i(2\r-\d)} \right)\right] + m_3 c_z s_z s_y,\nn\\
&&M_{\n\,13}=  m_1 \left[ -c_z s_z \left(c_x^2 s_y e^{2i\r} + s_x^2 c_y e^{2i\s}\right) +  c_z c_x s_x s_y \left(e^{i(2\r-\d)}- e^{i(2\s-\d)}\right) \right] + m_3 c_z s_z c_y,\nn\\
&& M_{\n\,22}= m_1 \left[ s_z^2 s_y^2 \left(c_x^2 e^{2i\r} + s_x^2 e^{2i\s}\right) + c_y^2 (s_x^2 e^{2i(\r-\d)} + c_x^2 e^{2i(\s-\d)})
+2 c_y c_x s_x s_z s_y \left(e^{i(2\r-\d)} - e^{i(2\s-\d)}\right) \right] + m_3 c_z^2 s_y^2, \nn\\
&& M_{\n\,33}= m_1 \left[s_z^2 c_y^2 \left(c_x^2 e^{2i\r} + s_x^2 e^{2i\s}\right) + s_y^2 \left(s_x^2 e^{2i(\r-\d)} + c_x^2 e^{2i(\s-\d)})
+ 2 c_y c_x s_x s_z s_y (-e^{i(2\r-\d)} + e^{i(2\s-\d)}\right) \right]
+ m_3 c_z^2 c_y^2,  \nn\\
&& M_{\n\, 23} = m_1 \left[ c_y s_y s_z^2 \left(c_x^2 e^{2i\r} + s_x^2 e^{2i\s}\right) +  s_z c_x s_x c_{2y}\left(e^{i(2\r-\d)}- e^{i(2\s-\d)}\right)
- c_y s_y \left(s_x^2 e^{i(2\r-\d)}+ c_x^2 e^{i(2\s-\d)}\right)\right] + m_3 s_y c_y c_z^2. \nn\\
\eea} and requiring them to correspond to the modified generic bipartite form (Eq. \ref{gen-bip-for}), so that the $\t_x$-dependence in the
mass matrix elements should drop out, we get $\r = \s$ and $\t_y=\pi/4$. Moreover, the last mass relation in Eq. (\ref{gen-deg-mass-rel}) for the $\mbox{\textsc{dntb}}$ bipartite form  leads to $\d=0$ since $M_{\n\,11}$ in Eq.(\ref{degenerate matrix elements})is $\d$-independent whereas $M_{\n_{22}}$ and $M_{\n_{23}}$ depend on $\d$. We conclude then that for the $S$-symmetry to be satisfied by the degenerate neutrino mass matrix, we need to have $\r=\s$ and $\d=0$.


\subsection{\large \bf The $Z_2$ and $Z_3$ subgroups in the DNTB pattern}
As was said earlier in the introduction, the symmetry $Z_2\times Z_3$ mentioned in \cite{matripar}
to characterize the degenerate mass spectrum case are special cases of the $U(1)$ S-symmetry we stated in the previous subsection. The
$Z_3$-symmetry corresponding to the DNTB pattern can be found by putting $\t=-\frac{2 \pi}{3}$ in Eq.(\ref{S-DNTB-sym}):
\bea
\label{S3-DNTB-sym}
S^{\mbox{\textsc{dntb}}}_3 \equiv S^{\mbox{\textsc{dntb}}}_{\t=-2\pi/3}&=&
\pmatrix{-\frac{3}{2} c_z^2 + 1 &  -\frac{\sqrt{2}}{4}c_z \left(\sqrt{3} - 3 s_z\right) & \frac{\sqrt{2}}{4}c_z \left(\sqrt{3} + 3 s_z\right)\cr \frac{\sqrt{2}}{4}c_z \left(\sqrt{3} + 3 s_z\right) & -\frac{1}{2}+\frac{3}{4} c_z^2
&- \frac{\sqrt{3}}{2}s_z + \frac{3}{4} c_z^2 \cr  -\frac{\sqrt{2}}{4}c_z \left(\sqrt{3} - 3 s_z\right) &  \frac{\sqrt{3}}{2}s_z + \frac{3}{4} c_z^2 & -\frac{1}{2}+\frac{3}{4} c_z^2}.
\eea
As for the $Z_2$-symmetry corresponding to the DNTB pattern \footnote{We denoted in the introduction this $Z_2$ subgroup of $U(1)$ by  $Z_2^\prime$. However, we shall drop the $^\prime$ mark, as it is clear from the context which $Z_2$ is meant.}, we get it by substituting $\t$ by $\pi$:
\bea
\label{S2-DNTB-sym}
S^{\mbox{\textsc{dntb}}}_2 \equiv S^{\mbox{\textsc{dntb}}}_{\t=\pi}&=&
\pmatrix{-c_{2z}&  \frac{1}{\sqrt{2}}\, s_{2z}  &  \frac{1}{\sqrt{2}}\, s_{2z} \cr \frac{1}{\sqrt{2}}\, s_{2z}  & -s_z^2
&   c_z^2 \cr  \frac{1}{\sqrt{2}}\, s_{2z}  &  c_z^2 & -s_z^2}.
 \eea

One can find similar equivalences to Eqs.(\ref{u1-gen-mass-equivalence},\ref{u1-gen-yuk-equivalence} and \ref{u1-gen-yuk-lef-equivalence})
corresponding to these subgroups.
For the $Z_2$-subgroup we find for all angles $\t_z$ the following:
\bea
\label{s2-gen-yuk-equivalence}
&&\left[   \left(S^{\mbox{\textsc{dntb}}}_2\right)^{\mbox{\textsc{t}}} \cdot f \cdot S^{\mbox{\textsc{dntb}}}_2=f  \right]  \Leftrightarrow  \left[\right.\;\exists\, a_1,a_2,c_1,c_2, c_3: \nonumber \\
&&f(1,1)=a_1,\; f(1,2)=a_2,\; f(1,3)= -a_2+2\,t_z^2\,c_1+\sqrt{2}\,t_z\,\left(c_3+c_2-a_1\right), \nonumber\\
&& f(2,1)=-\frac{1-3s_z^2}{c_z^2}\,c_1 +\sqrt{2}\,t_z\,\left(c_3+c_2-a_1\right),\; f(2,2)=c_3 +\sqrt{2}\,t_z\,\left(c_1-a_2\right),\nonumber\\
&&f(2,3)=\frac{1-3s_z^2}{c_z^2}\, \left(c_2+\sqrt{2}\,t_z\,c_1\right)+\sqrt{2}\,t_z\,\left[a_2+\sqrt{2}\,t_z\,\left(a_1-c_3\right)\right],
\nonumber\\
&& f(3,1)=c_1, f(3,2)=c_2, f(3,3)=c_3\left.\right].
\eea

In order to impose a symmetric matrix condition, which would be useful for a neutrino mass matrix, it suffices to make the following substitution
\be
c_2\rightarrow a_1-c_3+\frac{1}{\sqrt{2}\,t_z}\,a_2+\frac{1-3\,s_z^2}{\sqrt{2}\, c_z\, s_z}\,c_1,
 \ee
 leaving us with four free parameters that can be cast in the following form for all angles $\t_z$:
\bea
\label{s2-gen-mass-equivalence}
&&\left[   \left(M=M^{\mbox{\textsc{t}}}\right) \wedge \left(S^{\mbox{\textsc{dntb}}}_2\right)^{\mbox{\textsc{t}}} \cdot M \cdot S^{\mbox{\textsc{dntb}}}_2=M  \right] \Leftrightarrow  \left[\right.\; \exists\; A,B,C,D : \nonumber\\
&& M(1,1)= A-B+C,\; M(2,2)=A+C,\; M(2,3)=M(3,2)=D,\nonumber\\
&&M(1,2)=M(2,1)=\frac{-4\sqrt{2}\,s_{2z}\,s_z^3\, B  -\sqrt{2}\, c_z^3\,\left(10\,c_z^2 +\sqrt{2} \,s_z -8\right)\, C + 4\,\sqrt{2}\, c_z\, \left(1-3\,s_z^2\right)\, D }{4\,s_z\, \left(1-3\,s_z^2\right)^2},\nonumber\\
&& M(1,3) = M(3,1) = \frac{2\sqrt{2}s_z^2c_z B + c_z (5\sqrt{2}c_z^2 + s_z -4\sqrt{2}) C}{2s_z (1-3s_z^2)},\nonumber\\
&& M(3,3)=A-\frac{s^2_{2z}}{2(1-3s_z^2)^2}B-\frac{2s_z c_{2z}-14\sqrt{2}c_z^2+11\sqrt{2}c_z^4+4\sqrt{2}}{\sqrt{2}(1-3s_z^2)^2}C +\frac{2s_z^2}{1-3s_z^2}D\left.\right],
\eea
We note that if,
 \be
 D=\frac{c_z^2}{1-3s_z^2}B+\frac{-s_{2z}c_z+2\sqrt{2}c_z^2-\sqrt{2}}{s_z(1-3s_z^2)}C,
  \ee
 then we get exactly the modified generic tripartite form with three free parameters characterizing the NTB $Z_2^3$-symmetry obtained in \cite{ntb}. This means that if a symmetric matrix satisfies the form invariance with respect to $Z_2^3$-symmetry then it satisfies it for the $Z_2$-symmetry realized by $S_2^{\mbox{\textsc{dntb}}}$. This is clear due to the fact that $S_2^{\mbox{\textsc{dntb}}}$ is, up to a sign, just a factor of $Z_2^3$-symmetry, as can be seen trivially in the diagonalized basis. Moreover, if we put
in addition $C=0$ we get exactly the modified generic bipartite form with two free parameters (Eq. \ref{gen-bip-for}) characterized by the S-symmetry.
This is also evident since the latter bipartite form corresponds to a degenerate mass spectrum where $m_1=m_2$, which is a special case of the general mass spectrum for the modified tripartite form. As for the equivalence in Eq.(\ref{u1-gen-yuk-lef-equivalence}), corresponding to the
symmetry acting only from the left, we get exactly the same form in
both the $Z_2$-symmetry and the $U(1)$ S-symmetry.

We may think that we need to impose $Z_3 \times Z_2$ in order to characterize the modified generic bipartite form (Eq. \ref{gen-bip-for}) in line
with what was stated in \cite{matripar}. However, in accordance with our findings in \cite{u1}, we checked that imposing the $Z_3$-symmetry
in the degenerate mass spectrum case, which is represented by $S_3^{DNTB}$, leads alone precisely to the same equivalences in Eqs.~(\ref{u1-gen-mass-equivalence}, \ref{u1-gen-yuk-equivalence}, \ref{u1-gen-yuk-lef-equivalence}), so we can write for all angles $\t_z$ the following:
\bea \left[  (M=M^T) \wedge \left( \forall \t, (S^{DNTB}_\t)^T \cdot M \cdot  S^{DNTB}_\t = M \right)  \right]
& \Leftrightarrow & \left[  (M=M^T) \wedge \left( \forall \t, (S^{DNTB}_3)^T \cdot M \cdot  S^{DNTB}_3 = M \right)  \right] \label{u1-z3-equi-mass}
 \nonumber\\
\left[  \forall \t, (S^{DNTB}_\t)^T \cdot f \cdot  S^{DNTB}_\t = f   \right]
& \Leftrightarrow & \left[   (S^{DNTB}_3)^T \cdot f \cdot  S^{DNTB}_3 = f  \right] \label{u1-z3-equi-yuk} \nonumber
\\
\left[  \forall \t, (S^{DNTB}_\t)^T \cdot f = f   \right]
& \Leftrightarrow & \left[    (S^{DNTB}_3)^T \cdot f = f  \right]  \Leftrightarrow \left[    (S^{DNTB}_2)^T \cdot f = f  \right] \label{u1-z3-z2-equi-yuk-lef}
,\eea

Thus, the $Z_3$-symmetry and the $U(1)$ S-symmetry are phenomenologically equivalent regarding the form invariance, and the question arises as to what lies behind this fact. For this, we examine again what symmetries would characterize the form invariance formula (Eq. \ref{form}), in the diagonal basis, restricting it to the
2-dim subspace corresponding to the mass eigenvalues $m_1$ and $m_2$. Any `special' unitary matrix in two dimensions is represented by a rotation, so we have:
\bea
R^{\mbox{\textsc{t}}}\left(\t\right)\,\cdot\, \mbox{diag}\left(m_1,m_2\right) R\left(\t\right) = \mbox{diag}\left(m_1,m_2\right) &:& R\left(\t\right) = \pmatrix{c_\t & s_\t \cr -s_\t & c_\t}
= c_\t\, \mbox{I} + i\, s_\t \,\s_2,
\eea
where the identity matrix $I$ and the Pauli matrix $\s_2$ are given by
 \bea
 \mbox{I}=\mbox{diag}(1,1), && \s_2= \pmatrix{0 & -i \cr i & 0}.
 \eea
  This leads to:
\bea c_\t^2 \,\mbox{diag}(m_1,m_2) + \frac{i}{2}\, s_{2\t}\, \left[\mbox{diag}(m_1,m_2),\s_2\right] + s_\t^2\, \s_2\, \mbox{diag}(m_1,m_2)\, \s_2 &=& \mbox{diag}(m_1,m_2)
\label{form-diag-condition}\eea
We see directly here that for any fixed given angle $\t \neq \pi$, the relation in Eq.(\ref{form-diag-condition}) cannot be met unless we have a degenerate mass spectrum ($m_1=m_2$), so $Z_n$-symmetry, corresponding to $\t=\frac{2\pi}{n}$, leads, as long as $n \neq 2$, to a degenerate spectrum and thus to the $U(1)$-symmetry, and vice versa, whence the mentioned equivalence. It is clear also now, that the residual symmetry after  the breaking of $U(1)$ due to mass splitting $m_1 \neq m_2$ is the subgroup $Z_2$ corresponding to $\t = \pi$, the only value satisfying
 Eq.(\ref{form-diag-condition}) for a non-degenerate mass spectrum. This  equivalence between $Z_n$ and $U(1)$ regarding the form invariance should be contrasted with the case of regular n-polygons which are not $U(1)$-invariant under the whole set of rotations around their centers by arbitrary angles,
 although they are $Z_n$-symmetric, in that they stay unchanged when the rotation angle is a multiple of $\frac{2\pi}{n}$.


 \subsection{\large \bf The  DNTB$_0$ pattern}
In order to make definite conclusions and precise predictions for the phenomenological analyses in the following sections, we specify the results in this subsection to the experimentally best fit degenerate mass spectrum case of the NTB pattern, the DNTB$_0$ pattern characterized by,
\be
\t_z=\arcsin{\left(\frac{1}{\sqrt{50}}\right)},\; \t_x\, \mbox{undetermined},\; m_1=m_2.
\ee

The mixing matrix becomes:
\bea
V^{x}_0 &=& \frac{1}{10} \pmatrix{ 7\,\sqrt{2}\, c_x &  7\,\sqrt{2}\, s_x & \sqrt{2} \cr
- \left( c_x + 5\, \sqrt{2}\, s_x\right)
& - \left( s_x - 5\, \sqrt{2}\, c_x\right) & 7
\cr  - \left( c_x - 5\, \sqrt{2}\, s_x\right) & - \left( s_x + 5\, \sqrt{2}\, c_x\right)
& 7}.
\label{Vx0}
\eea
The modified special bipartite form is:
 \bea
\label{spe-bip-for}
M^{\mbox{\textsc{dntb}}_0}_{\n} &=& \pmatrix{A-B & \frac{7\sqrt{2}}{47} B & \frac{7\sqrt{2}}{47} B
\cr \frac{7\sqrt{2}}{47} B & A
&  \frac{49}{47} B\cr  \frac{7\sqrt{2}}{47} B &  \frac{49}{47} B & A},
\eea
and the eigen masses can be determined in terms of the bipartite form coefficients as:
\bea
\label{spe-deg-eigen-mass}
m_1 =m_2 =  A-\frac{49}{47}\,B, && A = \frac{51}{100}\, m_1 + \frac{49}{100}\, m_3
\nonumber, \\
m_3 = A+\frac{51}{47}\,B, && B = -\frac{47}{100}\, m_1 + \frac{47}{100}\, m_3 .
\eea
The $U(1)$ S$_0$-symmetry which characterizes this modified special bipartite form is given by:
\bea
\label{S-DNTB0-sym}
S^{\mbox{\textsc{dntb}}_0}_\t&=& \pmatrix{\frac{1}{50}\, \left(1+ 49\,c_\t\right)  & \frac{7\sqrt{2}}{100} \,\left(1- c_\t\right) + \frac{7}{10}\, s_\t &  \frac{7\sqrt{2}}{100}\, \left(1- c_\t\right) - \frac{7}{10}\, s_\t
\cr \frac{7\sqrt{2}}{100}\, \left(1- c_\t\right) - \frac{7}{10} s_\t & \frac{1}{100}\,               \left(49+ 51\,c_\t\right)
&  \frac{49}{100}\,\left(1- c_\t\right) + \frac{\sqrt{2}}{10}\, s_\t
\cr  \frac{7\sqrt{2}}{100}\,\left(1- c_\t\right) + \frac{7}{10}\, s_\t &
\frac{49}{100}\,\left(1- c_\t\right) - \frac{\sqrt{2}}{10}\, s_\t & \frac{1}{100}\,\left(49+ 51\,c_\t\right)}.
 \eea
The corresponding $Z_3$ and $Z_2$ symmetries are given by:
\bea
S^{\mbox{\textsc{dntb}}_0}_3 \equiv S^{\mbox{\textsc{dntb}}_0}_{\t=-2\pi/3}&=&
\pmatrix{-\frac{47}{100}  &  \frac{21\sqrt{2}}{200}-\frac{7\sqrt{3}}{20} & \frac{21\sqrt{2}}{200}+\frac{7\sqrt{3}}{20}
\cr \frac{21\sqrt{2}}{200}+\frac{7\sqrt{3}}{20} & \frac{47}{200}
& \frac{147}{200}- \frac{\sqrt{6}}{20}  \cr  \frac{21\sqrt{2}}{200}-\frac{7\sqrt{3}}{20} &  \frac{147}{200}+ \frac{\sqrt{6}}{20} & \frac{47}{200} },
\label{S3-DNTB0-sym}\\
S^{\mbox{\textsc{dntb}}_0}_2 \equiv S^{\mbox{\textsc{dntb}}_0}_{\t=\pi}&=&
\frac{1}{50}\pmatrix{-48  &  7\sqrt{2} & 7\sqrt{2}
\cr 7\sqrt{2} & -1& 49  \cr  7\sqrt{2} &  49 & -1 }.
\label{S2-DNTB0-sym}
\eea
Note that, as expected, the $\t_x$-dependence should not appear neither in the bipartite form, nor in the characterizing symmetry.

We have the corresponding equivalences:
\bea
\label{u1-spe-mass-equivalence}
&\left\{ \left(M=M^{\mbox{\textsc{t}}}\right)\wedge \left[\forall \t, \left(S^{\mbox{\textsc{dntb}}_0}_\t\right)^{\mbox{\textsc{t}}} \cdot M \cdot S^{\mbox{\textsc{dntb}}_0}_\t=M \right] \right\}
 \Leftrightarrow   \left\{ \left(M=M^{\mbox{\textsc{t}}}\right)\wedge \left[ \left(S^{\mbox{\textsc{dntb}}_0}_3\right)^{\mbox{\textsc{t}}} \cdot M \cdot S^{\mbox{\textsc{dntb}}_0}_3=M \right] \right\} &\nonumber \\& \Leftrightarrow {\small \left[\; \exists\; A,B: M = \pmatrix{A-B & \frac{7\sqrt{2}}{47} B & \frac{7\sqrt{2}}{47} B
\cr \frac{7\sqrt{2}}{47} B & A
&  \frac{49}{47} B\cr  \frac{7\sqrt{2}}{47} B &  \frac{49}{47} B & A}
\right]} &
\eea
and
\bea
\label{u1-spe-yuk-equivalence}
&\left[ \forall \t, \left(S^{\mbox{\textsc{dntb}}_0}_\t\right)^{\mbox{\textsc{t}}} \cdot f \cdot S^{\mbox{\textsc{dntb}}_0}_\t=f  \right]
  \Leftrightarrow \left[  \left(S^{\mbox{\textsc{dntb}}_0}_3\right)^{\mbox{\textsc{t}}} \cdot f \cdot S^{\mbox{\textsc{dntb}}_0}_3=f \right] & \nonumber \\& \Leftrightarrow   {\small \left[\; \exists\; A,B,C: f = \pmatrix{A-B+\frac{\sqrt{2}}{7}C & C & \frac{14\sqrt{2}}{47} B -\frac{51}{47} C
\cr\frac{14\sqrt{2}}{47} B -\frac{51}{47} C & A
&   B\cr  C &  \frac{51}{47} B - \frac{14\sqrt{2}}{47} C & A}
\right]}&
\eea
and
\bea
\label{u1-spe-yuk-lef-equivalence}
&\left[ \forall \t, \left(S^{\mbox{\textsc{dntb}}_0}_\t\right)^{\mbox{\textsc{t}}} \cdot f =f  \right]
  \Leftrightarrow \left[  \left(S^{DNTB_0}_3\right)^{\mbox{\textsc{t}}} \cdot f =f \right] \Leftrightarrow   {\small \left[\; \exists\; A,B,C: f = \pmatrix{\frac{\sqrt{2}A}{7} & \frac{\sqrt{2}B}{7} & \frac{\sqrt{2}C}{7}
\cr A & B & C
\cr  A &   B  & C}
\right]}&
\eea
One can also find the corresponding equivalences for the $Z_2$-symmetry for this special DNTB$_0$ pattern, and deduce that the resulting symmetric form, say,
contains the $Z_2^3$ modified special tripartite form attained in \cite{ntb} (Eq. \ref{gen-tri-for} with $\t_x=\arcsin(1/\sqrt{3}), \t_z=\arcsin(1/\sqrt{50})$), which
in turn includes the modified special bipartite form.



\section{{\large \bf The S$_0$-symmetry in the whole lepton sector}}
We impose now the S$_0$-symmetry in a setup involving also the charged leptons, since their LH components couple to the neutrinos, and any symmetry imposed on the latter should be met by the LH charged leptons as well.

 \subsection{\large \bf Model with many Higgs doublets}
\label{3.1}
 We follow here the model presented in \cite{matripar} and assume one heavy Higgs triplet $(\xi^{++},
\xi^+, \xi^0)$ and three scalar doublets $(\phi^0_i, \phi^-_i)$ playing the role of the standard model (SM) Higgs doublet.
\begin{eqnarray}
\label{lagrangian}
{\cal L}_Y = h_{ij} \left[\xi^0 \nu_i \nu_j - \xi^+ \left(\nu_i l_j + l_i
\nu_j\right)/ \sqrt 2 + \xi^{++} l_i l_j\right] + f_{ij}^k \left(l_i \phi^0_j -
\nu_i \phi^-_j\right) l^c_k + h.c.,
\end{eqnarray}
where under the S$_0$-symmetry the fields are transformed as:
\begin{eqnarray}
&& \left(\nu,l\right)_i \to \left(S_\th\right)_{ij} \left(\nu,l\right)_j, ~~~ l^c_k \to l^c_k, \\
&& \left(\phi^0,\phi^-\right)_i \to \left(S_\th\right)_{ij} \left(\phi^0,\phi^-\right)_j, ~~~
\left(\xi^{++}, \xi^+, \xi^0\right) \to \left(\xi^{++}, \xi^+, \xi^0\right).
\end{eqnarray}
Invariance of the Lagrangian means we have:
\bea
\label{h}
S_\th^T\; h\; S_\th &=& h,
\\
\label{f}
S_\th^T\; f^k \; S_\th &=& f^k.
\eea
This Lagrangian has a global symmetry $U(1)_L\bigotimes S_{0}$,
where $U(1)_L$ is the total lepton number symmetry, where we assign a zero lepton number to the doublets $\phi_i$ and a two lepton number for the heavy triplet $\xi$.
We add now a soft symmetry breaking `$\mu$'-term \begin{eqnarray}
\delta {\cal L_Y} =  \frac{\mu_{ij}}{2}\phi_i^T \xi^\dagger i\tau_2 \phi_j +h.c.
\end{eqnarray}
where the symmetric matrix $\mu_{ij}$ is not proportional, neither to the identity nor to the form dictated by $S_\t$, so that both the $U(1)_L$-symmetry and the $S_0$-symmetry are
broken explicitly. This term is introduced to avoid a Goldstone Boson (GB) associated with the spontaneous breaking of total lepton number (called Majoron) \cite{SV}. This will  have the effect of inducing a mass to the would-be GB of the order the mass of the Higgs triplet, and hence avoiding an invisible decay of the  Z gauge boson. In addition, this tadpole term will generate, upon minimization of the potential with respect to the neutral component of the triplet scalar $\xi^0$  a vacuum expectation value (vev) given by: \begin{eqnarray}
<\xi^0> = \frac{-\mu_{ij}  v_i v_j }{M^2_{\xi}},
\end{eqnarray} which can be small in the electron volt range, in line with a naturally tiny neutrino mass, for $\mu_{ij} \sim M_{\xi} \mbox{(the triplet mass)} \sim 10^{12}\,\mbox{GeV}$ \cite{W,ma03} \footnote{We could in principle choose $M_{\xi} \sim \mbox{TeV}$, which makes the triplet field accessible to colliders, but that will require choosing the coupling $h_{ij}$ unnaturally small ($\sim 10^{-10}$) unless one chooses $\mu \sim \mbox{eV}$.}. Furthermore, the tree level correction to the electroweak (EW) $\rho$ parameter  is $\sim \left(<\xi^0>/{v_{EW}}\right)^2 \sim \left(m_{\nu}/{v_{EW}}\right)^2$ which is negligible.

Moreover, This `$\mu$-term in the scalar sector will not destabilize the structure of the neutrino mass matrix, since the latter is dictated by how the leptonic fields transform under $S_\theta$.  However, there will be Yukawa-like interactions between the neutrinos and the pseudo-GB, and between the charged leptons and the electrically  charged components of the triplet field. But since all these scalar fields (including the pseudo-GB) are much heavier than the TeV scale, they do not have  observable effects on the fermion sector of the model. Said differently, the components of the triplet field will practically decouple from the low energy spectrum  and one is left only with the SM degrees of freedom plus effective higher dimensional operators suppressed by the  mass square of the  triplet (and a correction to the Higgs self-coupling) of the form:
$$
\mu \frac{LL \Phi \Phi}{M^2_\xi}
$$
which after the EW symmetry breaking gives $m_{\nu} \sim  \mu v^2/{M^2_{\xi}} \ll v.$

The equivalences (Eqs. \ref{u1-spe-mass-equivalence} and \ref{u1-spe-yuk-equivalence}) for the symmetric matrix $h_{ij}$ and the not-necessarily
 symmetric Yukawa couplings $f^k_{ij}$ leads to the forms:
 \bea
 \label{mod-1-cha-lep-h-f}
h = \pmatrix{A-B & \frac{7\sqrt{2}}{47} B & \frac{7\sqrt{2}}{47} B
\cr \frac{7\sqrt{2}}{47} B & A
&  \frac{49}{47} B\cr  \frac{7\sqrt{2}}{47} B &  \frac{49}{47} B & A}, && f^k = \pmatrix{a_k-b_k+\frac{\sqrt{2}}{7}c_k & c_k & \frac{14\sqrt{2}}{47} b_k -\frac{51}{47} c_k
\cr\frac{14\sqrt{2}}{47} b_k -\frac{51}{47} c_k & a_k
&   b_k\cr  c_k &  \frac{51}{47} b_k - \frac{14\sqrt{2}}{47} c_k & a_k}.
\eea
The neutrino mass matrix, when $\xi^{0}$ gets a vev, is
\bea \label{mod-1-neut-mass} (M_\n)_{ij} &=& \langle \xi^0 \rangle h_{ij},
\eea
which expresses the translation of the $S_0$-symmetry from the symmetric Yuakawa couplings $h_{ij}$ to the neutrino mass matrix $M_{\n_{ij}}$.
As to the charged leptons, the Yaukawa term $(f^k_{ij} l_i l^c_k \phi^0_j)$ leads, when the SM-like Higgs fields take vevs $(v_j=\langle \phi^0_j \rangle)$, to the mass matrix
\be
\left(M_l\right)_{ik} = f^k_{ij} v_j,
\ee
 The Yukawa couplings can be arranged so that to bring, after suitably rotating the charged RH singlet lepton $l^c$, the charged lepton mass matrix
  into its form in the flavor basis. For example, if $v_{1,2} << v_3$ we have:
\begin{eqnarray}
\label{mod-1-clm}
 M_l &=& v_3 \pmatrix{ A'_1  & A'_2 & A'_3 \cr B'_1 &
B'_2 & B'_3 \cr C'_1 & C'_2 & C'_3 },
\end{eqnarray}
where
\bea
&A'_i = \frac{14\sqrt{2}}{47} b_i - \frac{51}{47} c_i \;\;\;\;\;\;,\;\;\;\;\;\; B'_i = b_i \;\;\;\;\;\;,\;\;\;\;\;\;
C'_i = a_i &
\eea
In \cite{ntb}, a charged lepton matrix of precisely the same form was shown to represent the lepton mass matrix in the flavor basis with the right charged lepton mass hierarchies, assuming
just the ratios of the magnitudes of the vectors comparable to the lepton mass ratios.


 \subsection{\large \bf Model with many heavy SM-singlets}
  The model with many Higgs doublets induces dangerous flavor changing neutral currents \cite{BjorWein77}. For this, we might think
  of keeping just one SM-Higgs doublet $\Phi$ but at the expense of adding three heavy SM-singlet scalars $\D_i$ transforming non-trivially under
  the $S_0$-symmetry. Again, we assume the SM Higgs and the charged RH leptons $l^c_j$  to be
singlets under the $S_0$-symmetry, whereas the lepton
 LH doublets transform component-wise faithfully:
\begin{equation}
 L_i
\rightarrow S^{\mbox{\textsc{dntb}}_0}_{ij}\;L_j,
\end{equation}
with $i,j =1,2,3$. Then, the invariance of the SM term
  \bea
\label{L1}
{\cal{L}}_1 &=& Y_{ij} \overline{L}_i \Phi  l^c_j,
\eea
under $S_0$-symmetry leads via (Eq. \ref{u1-spe-yuk-lef-equivalence}) to the form:
\bea
\label{mod-2-cha-lep-yuk-lef}
Y_{ij} &\sim & \pmatrix{\frac{\sqrt{2}}{7} A & \frac{\sqrt{2}}{7} B & \frac{\sqrt{2}}{7} C \cr
A&B&C \cr A&B&C
 }.
 \eea
We see here that this term leads, when $\Phi^0$ gets a vev, to a  charged lepton squared mass matrix proportional to
\[Y\cdot Y^\dagger \sim \frac{\left(|A|^2 + |B|^2 +|C|^2\right)}{7} \pmatrix{\frac{2}{7} & 1& 1\cr 1& 7 & 7 \cr 1 & 7 & 7 },\]
with two zero eigenvalues, so we can not produce the charged lepton mass spectrum by this term. Moreover, $YY^\dagger$ has eigenvectors
proportional to $(1,-\frac{\sqrt{2}}{7},0),(0,-1,1),(\frac{\sqrt{2}}{7},1,1)$, which means that $YY^\dagger$ is not diagonalized trivially, and so we are not in the flavor basis, which would destroy the predictions of the ${\mbox{\textsc{dntb}}_0}$ pattern. We note here that had we really
taken the original full symmetry of the model ($F=U(1) \times Z_2 \times Z_2 = \langle U(1), Z_2^3 \rangle$) then the equation $F\cdot Y = Y$ can not be met
for all elements in $F$ unless $Y=0$.

The additional heavy SM-singlet scalar fields $\D_i$ help in resolving these inconveniences. We assume them to transform under $S_0$-symmetry as
\begin{equation}
 \Delta_i \rightarrow S^{\mbox{\textsc{dntb}}_0}_{i j}\Delta_{j}.
\end{equation} and that they are coupled to the lepton LH doublets through a non-renormalizable dimension-5 operator
\begin{eqnarray}
{\cal{L}}_2 &=& \frac{f_{ikr}}{\Lambda} \overline{L}_i \Phi \Delta_k  l^c_r ,
\end{eqnarray}
where $\Lambda$ is a heavy mass scale. Invariance of ${\cal{L}}_2$  under $S_0$-symmetry leads to
\bea \label{f_r}
\left(S^{\mbox{\textsc{dntb}}_0}\right)^{\mbox{\textsc{t}}} \, f_r \, S^{\mbox{\textsc{dntb}}_0} =f_r , \eea
where $f_r$, for fixed $r$, is the matrix whose ($i,j$) entry is $f_{i j r}$.
The equivalence (Eq. \ref{u1-spe-yuk-equivalence}) leads to:
\begin{eqnarray}
\label{cly} f_r = \pmatrix{A_r-B_r+\frac{\sqrt{2}}{7}C_r & C_r & \frac{14\sqrt{2}}{47} B_r -\frac{51}{47} C_r
\cr\frac{14\sqrt{2}}{47} B_r -\frac{51}{47} C_r & A_r
&   B_r\cr  C_r &  \frac{51}{47} B_r - \frac{14\sqrt{2}}{47} C_r & A_r} .\end{eqnarray}
 When $\Delta_k$ and $\phi^\circ$ take the vevs
$\delta_k$ and $v$ respectively, then we get the charged lepton mass
 matrix:
\begin{equation}
\left( {\cal{M}}_l \right)_{ir} = \frac{v f_{ikr}}{\Lambda}\delta_k.
\end{equation}
Again, one can arrange the vevs and the Yukawa couplings such that ${\cal{M}}_l$ after suitably rotating the flavor- and SM-singlets $l^c_j$
is the charged lepton mass matrix in the flavor basis. For example, if  $\delta_1,\delta_2 \ll \delta_3$ we get:
\begin{eqnarray}\label{mod-2-clm}
 M_l &=& \frac{v\delta_3}{\Lambda} \left ( \matrix{ A'_1  & A'_2 & A'_3 \cr B'_1 &
B'_2 & B'_3 \cr C'_1 & C'_2 & C'_3 \cr} \right ),
\end{eqnarray}
where
\bea
A'_i = (\frac{14\sqrt{2}}{47} B^i-\frac{51}{47}C^i), & B'_i = B^i, & C'_i = A^i.
\eea
The same diagonalization procedure mentioned in the last subsection can be applied here to show that ${\cal{M}}_l$ can be seen, to a good approximation, as the charged lepton mass matrix with the correct mass hierarchies in the flavor basis.


\subsection{\large \bf The conserved current associated with the S-symmetry}
One can determine the conserved current and charge corresponding to the continuous $U(1)$-symmetry. In order to stress the generality
of the treatment, we shall discuss the S-symmetry with generic values  of $\t_z$. Let us for illustration purposes consider the neutrino part where
the relevant term for computing the current is the kinetic energy one (sum is understood over the flavor index $k$):
\bea
\label{kinetic energy }
K_\n &=& i \bar{\nu}_k \gamma^\mu \partial_\mu \nu_k.  \eea
The current associated to the S-symmetry (Eq. \ref{S-DNTB-sym}) is
given by: \bea J^\mu_\n \equiv -i\,\frac{\partial{K_\n}}{\partial(\partial_\mu \n_j)} T_{jk} \n_k =  T_{jk} \bar{\n}_j \gamma^\mu \n_k
\label{current}
\eea
where $T_{ij}$ is the generator of the S-symmetry :
\bea \label{generator}
T &=& i
    \pmatrix {0 & {c_z\over \sqrt{2}} & {-c_z\over \sqrt{2}} \cr
    {-c_z\over \sqrt{2}} & 0 & s_z \cr
     {c_z\over \sqrt{2}} & -s_z & 0},
     \eea
satisfying
\be
S^{\mbox{\textsc{dntb}}}_{\d\t} \approx \mbox{I} -i\, \d \t\; T.
\ee

Moreover, since $S_\t^{\mbox{\textsc{dntb}}}$ is a 3-dimensional representation of the commutative $U(1)$ group whose irreducible representations (irreps)  are 1-dimensioanl, one must
be able to reduce $S_\t^{\mbox{\textsc{dntb}}}$ into three 1-dimensional irreps obtained by diagonalizing $S_\t^{\mbox{\textsc{dntb}}}$ as follows:
\bea \label{eigenfield}
S^{\mbox{\textsc{dntb}}}_\t &=& L  \pmatrix{1 & 0 & 0 \cr 0 & e^{-i\t} & 0 \cr 0 & 0
& e^{i\t}} L^\dagger ,
\eea
\bea
\label{ L expression} L &=& \pmatrix{s_z & \frac{-c_z(i+s_z)}{\sqrt{2(1+s_z^2)}} & \frac{-c_z(-i+s_z)}{\sqrt{2(1+s_z^2)}}
\cr \frac{c_z}{\sqrt{2}} & \frac{-c^2_z+2i s_z}{2\sqrt{(1+s_z^2)}} & \frac{-c^2_z-2i s_z}{2\sqrt{(1+s_z^2)}}
\cr \frac{c_z}{\sqrt{2}} & \frac{\sqrt{1+s_z^2}}{2}
& \frac{\sqrt{1+s_z^2}}{2}} = \pmatrix{V_0 & V_- & V_+}.
\eea
The `neutrino' eigenvectors $(V_0,V_-,V_+)$, with expressions in terms of the flavor or `gauge' states given by the columns of $L$, are the neutrino fields with definite $S^{DNTB}_\t$-charges equaling
respectively to ($0,-1,+1$). Inverting now, to express the neutrino gauge states $\n_i$ in terms of ($V_0,V_-,V_+$), and
substituting in Eq.(\ref{current}) we get:
\bea
\label{neutcurrent}
J^\m_\n &=& \left( 0\, \bar{V}_0\,\gamma^\mu\, V_0 - 1\, \bar{V}_-\,\gamma^\mu\, V_-
+ 1\, \bar{V}_+\, \gamma^\mu \,V_+\right),
\eea
which expresses explicitly the conserved current in terms of the S-charge eigenstates. This current corresponds to a global non-gauged continuous symmetry, similar to the
$U(1)$ baryon number conservation in the SM.

Using now $V^{xz}$ (Eq. \ref{Vxz}) to go from the neutrino
`gauge' states $\n^{\bf g}= \left(\n_e\,  \n_\mu\,  \n_\tau\right)^{\mbox{\textsc{t}}}$ to neutrino `mass' states $\n^{\bf m} = \left(\n^m_1\, \n^m_2\, \n^m_3\right)^{\mbox{\textsc{t}}}$, we express the definite $S^{\mbox{\textsc{dntb}}}_\t$-charge neutrino fields
in terms of the mass eigenstates:
\be
{\bf V}\equiv \left(V_0\; V_-\;  V_+\right)^{\mbox{\textsc{t}}} =L^T \cdot \n^{\bf g} = L^T \cdot (V^{xz})^* \cdot \n^{\bf m},
\ee
 which gives:
\bea
V_0 & = & \nu^m_3, \nn\\
V_- & =& a (\n^m_1 - i \n^m_2), \nn\\
V_- & =& a^* (\n^m_1 + i \n^m_2),\;\;\mbox{where}\; a = -\frac{i+s_z}{\sqrt{2(1+s_z^2)}}e^{ix}.
\label{chnum}
\eea
Although the expressions involve $\t_x$, however and as expected, this phase has no `physical'
content in the degenerate mass spectrum case, since the particular combination of mass eigen-states in Eq.(\ref{chnum}) never
mix under free time evolution provided $\nu^m_1$ and $\nu^m_2$ have degenerate mass, which is the case when the S-symmetry is exact. The same conclusion still holds if one think of the
underlying symmetry, in the degenerate two masses case, as  $Z_3 \times Z_2$, due to the compatibility of
both $S^{\mbox{\textsc{dntb}}}_\t$ and $Z_3\times Z_2$ in that they all commute and have common eigen-states. In fact, as we have seen earlier, even in the non-degenerate spectrum case, the mass eigensates ($\n^m_1,\n^m_2,\n^m_3 $) are the eigenvectors of the residual $Z_2$-symmetry, to which we can attribute `conserved' charges equal respectively to ($-1,-1,1$).



\section{The DNTB$_0$ neutrino mass matrix in seesaw scenarios }

We saw in Eqs. (\ref{nor-hier},\ref{inv-hier} and \ref{deg-hier}) that the modified bipartite form can
explain all sorts of neutrino mass hierarchies. In the next subsection, we shall be more specific on the origin of the coefficients
of the bipartite form through invoking type-I seesaw scenarios.

\subsection{\large \bf Type-I seesaw scenario in the DNTB pattern}
The effective light LH neutrino mass matrix is generated through the seesaw formula:
\be
M_\nu = - M_\nu^D \cdot  M_R^{-1} \cdot  \left(M_\nu^D\right)^{\mbox{\textsc{t}}},
\label{seesaw}
\ee
where $M_R$ is the heavy Majorana RH neutrinos mass matrix, whereas the Dirac neutrino mass matrix comes from the Yukawa term:
\begin{eqnarray}
g_{ij}\; \overline{L}_i\;
\tilde{\Phi}\; \nu_{Rj},
\end{eqnarray}
with $\tilde{\Phi} = i \tau_2 \Phi^*$. Again, for generality, we will treat in this subsection the S-symmetry corresponding to the DNTB pattern with generic values of $\t_z$ since the results are not specific to any particular value of it. We assume the RH neutrinos transforming under $S$-symmetry as
\begin{eqnarray} \label{rtransform}
\nu_{Rj} \rightarrow\; S_{j\gamma}\, \nu_{R\gamma},
\end{eqnarray}
Then, the invariance of the Lagrangian under S-symmetry leads to
\begin{eqnarray}
S^{\mbox{\textsc{t}}} \cdot g \cdot S=g .
\end{eqnarray}
The equivalence in Eq.(\ref{u1-gen-mass-equivalence}) leads, when $\tilde{\Phi}$ takes a vev $v$, to the following Dirac mass matrix:
\bea
\label{mD-seesaw1}
M^D_\n &=& v \pmatrix{A_D-B_D+\sqrt{2}t_z C_D & C_D
& \frac{\sqrt{2} s_{2z}}{1-3s_z^2} B_D - \frac{1+ s^2_z}{1-3s_z^2} C_D
\cr \frac{\sqrt{2} s_{2z}}{1-3s_z^2} B_D - \frac{1+ s^2_z}{1-3s_z^2} C_D & A_D
&  B_D\cr  C_D &    \frac{1+ s^2_z}{1-3s_z^2} B_D - \frac{\sqrt{2} s_{2z}}{1-3s_z^2} C_D& A_D}.
\eea
The invariance under S-symmetry of the term $\frac{1}{2}\, \nu^{\mbox{\textsc{t}}}_{iR}\, C \left( M_R \right)_{ij} \nu_{jR}$ ($C$ is the charge conjugation matrix)
would impose the modified generic bipartite form for the symmetric Majorana RH neutrino mass matrix (Eq. \ref{u1-gen-mass-equivalence}):
\begin{eqnarray}
\label{mR-seesaw1}
 M_R = \Lambda_R \pmatrix{A_R-B_R & \frac{\sqrt{2} s_z c_z}{1-3s_z^2} B_R & \frac{\sqrt{2} s_z c_z}{1-3s_z^2} B_R
\cr \frac{\sqrt{2} s_z c_z}{1-3s_z^2} B_R & A_R
&  \frac{c_z^2}{1-3s_z^2} B_R\cr  \frac{\sqrt{2} s_z c_z}{1-3s_z^2} B_R &  \frac{c_z^2}{1-3s_z^2} B_R & A_R}.
 \end{eqnarray}
where $\Lambda_R$ is a high scale characterizing the seesaw mechanism.

Applying the seesaw formula (Eq. \ref{seesaw}) we get the same form characterizing the DNTB pattern:
\begin{eqnarray}
\label{nRm}
 M_\n = -\frac{v^2}{\Lambda}\pmatrix{A_\n-B_\n & \frac{\sqrt{2} s_z c_z}{1-3s_z^2} B_\n & \frac{\sqrt{2} s_z c_z}{1-3s_z^2} B_\n
\cr \frac{\sqrt{2} s_z c_z}{1-3s_z^2} B_\n & A_\n
&  \frac{c_z^2}{1-3s_z^2} B_\n\cr  \frac{\sqrt{2} s_z c_z}{1-3s_z^2} B_\n &  \frac{c_z^2}{1-3s_z^2} B_\n & A_\n},
 \end{eqnarray}
where the bipartite form coefficients $A_\n, \; B_\n$ are given in terms of those characterizing the Dirac and Majorana mass matrices as follows.
\bea
\label{A-coefficients}
A_\n &=& \frac{A_{\n_1} A_R + A_{\n_2} B_R}{(1-3s_z^2)\left[ (1-3s_z^2)A_R-c_z^2B_R\right]\left[  (1-3s_z^2)A_R+(1+s_z^2)B_R\right]},\nn\\
A_{\n_1} &=& (1-3s_z^2)^3A_D^2+(1-3s_z^2)(1+s_z^2)^2C_D^2-2\sqrt{2}s_{2z}(1+s_z^2)(1-3s_z^2)B_DC_D
+(1+s_z^2)^2(1-3s_z^2)B^2_D, \nn \\
A_{\n_2}&=&  2s_z^2(1-3s_z^2)^2A_D^2+\sqrt{2}s_{2z}(1+s_z^2)(1-3s_z^2)A_DC_D-2c_z^2(1+s_z^2)(1-3s_z^2)A_DB_D + \nn \\ &&
c_z^2 (1+s_z^2)^2 C_D^2 -\sqrt{2} s_{2z}(1+s_z^2)^2 B_DC_D + 2 s_z^2 (1+s_z^2)^2 B_D^2,
\eea
\bea
\label{B-coefficients}
B_\n &=& \frac{B_{\n_1} A_R + B_{\n_2} B_R}{-c_z^2\left[  (1-3s_z^2)A_R-c_z^2B_R\right]\left[  (1-3s_z^2)A_R+(1+s_z^2)B_R\right]},\nn\\
B_{\n_1} &=& \sqrt{2} s_{2z}(1-3s_z^2)^2A_D C_D-2c_z^2(1-3s_z^2)^2A_DB_D+(1+s_z^2)(1-3s_z^2)^2C^2_D-\sqrt{2}s_{2z}(1-3s_z^2)B_DC_D, \nn \\
B_{\n_2}&=&  c_z^2(1-3s_z^2)^2A_D^2+c_z^2(1+s_z^2)^2C^2_D-4\sqrt{2}s_zc_z^3(1+s_z^2)B_DC_D -c_z^2 (1+s_z^2)B_D^2.
\eea

All types of neutrino mass hierarchies can be accommodated according to relations in Eqs.~(\ref{nor-hier}-\ref{deg-hier}) which in turn impose constraints on Dirac and RH Majorana neutrino mass matrices as follows;
\begin{itemize}
\item{ Normal hierarchy}: with
\begin{eqnarray} A_D \simeq
B_D,\; C_D \ll B_D\; (A_D),\; A_R \simeq
B_R,
\end{eqnarray}
we get,  for most values of $\t_z$ in the experimentally acceptable range ($[6.29^o,11.68^o]$), the following
\begin{eqnarray}
A_\n \simeq \frac{A_D^2}{A_R} \simeq B_\n &\Rightarrow& A_\n \simeq B_\n.
\end{eqnarray}
\item{ Inverted hierarchy}: with
\begin{eqnarray} A_D \simeq
- B_D,\; C_D \ll B_D\;(A_D) ,\; A_R \simeq
-B_R,
\end{eqnarray}
we get,  for most acceptable values of $\t_z$, the following
\begin{eqnarray}
A_\n \simeq \frac{A_D^2}{A_R} \simeq - B_\n &\Rightarrow& A_\n \simeq - B_\n.
\end{eqnarray}
\item{ Degenerate case}: with
\begin{eqnarray} A_D \gg B_D \gg C_D, \; A_R \gg B_R,
\end{eqnarray}
we get,  for most acceptable values of $\t_z$, the following
\begin{eqnarray}
A_\n \simeq \frac{A_D^2}{A_R},\; B_\n \simeq 2\frac{B_DA_D}{A_R} - A_D^2 \frac{B_R}{A_R^2} &\Rightarrow& A_\n \gg  B_\n.
\end{eqnarray}
\end{itemize}

The RH neutrino mass term violates lepton number by two units, and could lead to lepton asymmetry. The produced asymmetry due to the out of equilibrium decay of the lightest RH  neutrino to SM particles is given by \cite{FY}:
\begin{eqnarray}
\epsilon \simeq \frac{3}{16\pi
v^2}\frac{1}{\left(\tilde{M}_D^{\dagger}\tilde{M}_D\right)_{11}}\sum_{j=2,3}
\mbox{Im}\left\{\left[(\tilde{M}_D^{\dagger}\tilde{M}_D)_{1j}\right]^2\right\}\frac{M_{R1}}{M_{Rj}},
\end{eqnarray}
where $M_{Ri},\; i=1\cdots 3$ are the masses for RH
neutrinos, and $\tilde{M}_D$ is the Dirac neutrino mass
matrix in the basis where the Majorana RH neutrino mass
matrix $M_{R}$ is diagonal.
Since the RH neutrino mass matrix (Eq. \ref{nRm}) has the DNTB modified generic bipartite form, then it is diagonalized by $V^{xz}$ (Eq. \ref{Vxz}). Thus under $\n_R \rightarrow \left(V^{xz}\right)^*\; \n_R$ we have $M^D \rightarrow M^D\; \left(V^{xz}\right)^*$. We still have
freedom in multiplying the diagonalizing unitary matrix $V^{xz}$ by diagonal phases $F = \mbox{diag}\left(e^{i \a_1},\; e^{i \a_2},\;e^{i \a_3} \right)$ adjusted  normally such that
the phases of the spectrum of $M_{R}$ disappear. Namely, these phases vanish if we choose
\bea
\left( \a_1, \a_2, \a_3 \right) &=& \frac{1}{2}\arg \left( A_R-\frac{c_z^2}{1-3s_z^2}B_R,\;
A_R-\frac{c_z^2}{1-3s_z^2}B_R,\; A_R+\frac{1+s_z^2}{1-3s_z^2}B_R
\right).
\eea
Thus, we have $\tilde{M}_D = M^D \cdot \left(V^{xz}\right)^* \cdot F^*$, so we can write
\bea
\tilde{M}_{D}^{\dagger} \cdot \tilde{M}_{D} &=& F^{\mbox{\textsc{t}}}\cdot \left(V^{xz}\right)^{\mbox{\textsc{t}}}\cdot M^{D \dagger}  \cdot M^{D} \cdot \left(V^{xz}\right)^* \cdot F^*.
\eea
The required entries of $\tilde{M}_{D}^{\dagger} \cdot \tilde{M}_{D}$ for calculating the asymmetry are,
\begin{eqnarray}
(\tilde{M}_\nu^{D\dagger}\tilde{M}_\nu^D)_{11} &=&
 \left|A_D\right|^2 + {1\over 4-12\,c_z^2 + 9\, c_z^4}\,\left[\right.\left(4\,s_z^2 +c_z^4\right)\left|B_D\right|^2 + \left(2\,c_z^2\,(1+s_z^2)\right)\,\left|C_D\right|^2 \nn\\
 &&-\sqrt{2}\,s_z\,c_z\,(c_z^2 +2)\,\left(C_D\; B^*_D  + C^*_D\; B_D\right)  + (2\,c_z^2 -3\, c_z^4)\,\left(A_D\; B^*_D +A^*_D\; B_D\right)\nn\\
 && +\sqrt{2}\,s_z\,c_z\,(3\,c_z^2 -2)\,\left(A_D\; C^*_D +A^*_D\; C_D\right)\left.\right]
 ,\nonumber \\
(\tilde{M}_\nu^{D\dagger}\tilde{M}_\nu^D)_{12} &=&
{1\over 1 - 3\,s_z^2}\left[ 2\,s_z\,\left(B^*_D\; A_D - B_D\; A^*_D\right)
+ \sqrt{2}\,c_z\,\left(C^*_D\; B_D - C_D\; B^*_D\right) + \sqrt{2}\,c_z\,\left(A^*_D\; C_D - A_D\; C^*_D\right)\right],\nonumber \\
(\tilde{M}_\nu^{D\dagger}\tilde{M}_\nu^D)_{13} &=& 0,
\end{eqnarray}
which leads to a vanishing lepton asymmetry since the entries $(1,1)$ and $(1,2)$ are respectively real and pure imaginary, whereas the entry $(1,3)$ assumes the value zero. Thus we are tempted to look for other phenomenologically motivated venues producing enough lepton asymmetry in the context of Type-II seesaw mechanism.


\subsection{\large \bf Type-II seesaw scenario in the DNTB$_0$ pattern}
 Type-II seesaw scenario can solely accommodate an enough lepton/baryogenesis for the observed baryon/photon density in the universe. For this, we need to do some numerical estimations, that is why we call forth the
special value $\t_z \sim 8^o$ of the DNTB$_0$ pattern in this section.
As in \cite{ntb}, we introduce two SM triplet fields $\Sigma_A$, $A=1,2$ which are singlet
under the S-symmetry. The Lagrangian part relevant for the neutrino mass matrix is:
\begin{equation}
{\cal{L} } = \lambda_{\alpha\beta}^{A}\, L_\alpha^T\, C\, \Sigma_A\, i\,\tau_2\, L_\beta +
{\cal{L}}(H,\Sigma_A)+h.c.
\end{equation}
where $A=1,2$ and
\begin{eqnarray}
{\cal{L}}(H,\Sigma_A) &=& \mu_H^2 H^\dagger H + \frac{\lambda_H}{2} {(H^\dagger H)}^2+
M_A\, \mbox{Tr}\left(\Sigma_A^\dagger \Sigma_A \right)+
\frac{\lambda_{\Sigma_A}}{2} \left[\mbox{Tr}\left( \Sigma_A^\dagger \Sigma_A\right)\right]^2
+ \\\nonumber && \lambda_{H\Sigma_A} (H^\dagger H) \mbox{Tr}\left( \Sigma^\dagger_A \Sigma_A\right)
+
{\mu_A H^T \Sigma_A^\dagger i\tau_2 H +h.c.}
\end{eqnarray}
where $H$ and $\Sigma_A$ are written as
\bea
H=\pmatrix{\phi^+ \cr \phi^0}, &&  \Sigma_A = \left ( \matrix{ \frac{\Sigma^+}{\sqrt{2}}  &
\Sigma^0 \cr \Sigma^{++} & -\frac{\Sigma^+}{\sqrt{2}}\cr} \right )_A.
\eea

The neutrino mass matrix due to the exchange of the two triplets, $\Sigma_1$ and $\Sigma_2$, is
\begin{equation} \label{mass}
(M_\nu)_{\alpha\beta}\simeq v^2 \left[\lambda^1_{\alpha\beta} \frac{\mu_1}{M^2_{\Sigma_1}} +
\lambda^2_{\alpha\beta}\frac{\mu_2}{M^2_{\Sigma_2}}\right]
\end{equation}
where $M_{\Sigma_i}$ is the mass of the neutral component $\Sigma_i^0$ of the triplet $\Sigma_i ,i=1,2$.

Some remarks are in order here. First, the two matrices $\lam^A$, and thus also the neutrino mass matrix in Eq. \ref{mass}, have
the modified special bipartite form because of the Lagrangian invariance under the $S_0$-symmetry:
\bea
 \lam^a = \pmatrix{A_a-B_a & \frac{7\sqrt{2}}{47} B_a & \frac{7\sqrt{2}}{47} B_a
\cr \frac{7\sqrt{2}}{47} B^a & A^a
&  \frac{49}{47} B\cr  \frac{7\sqrt{2}}{47} B^a &  \frac{49}{47} B^a & A^a},\;a=1,2
\eea
Hence, all kinds of neutrino mass hierarchies can be generated. Second, the same remarks mentioned in subsection \ref{3.1} about the $\mu_A$-term apply here concerning the generation of small neutrino mass and the integrating out of the heavy triplet and the pseudo-GB. Third, the flavor changing  neutral current due to the triplet is highly
suppressed because of the large value of its mass scale.

We follow now the same steps carried out in \cite{ntb,u1,z2} discussing how to generate baryon asymmetry from leptogenesis using the sphaleron interaction \cite{KRS}. We note in passing that the would-be GB, which is the imaginary part of the neutral component of the triplet, would contribute to the leptogenesis as any component of this field. The reason
 behind this is that at temperatures much larger than the EW scale, the vev of the triplet is equal to zero since it is  triggered by the spontaneous  breaking of the SM gauge symmetry (which at these temperatures is unbroken). That is the $SU(2)\times U(1)$  gauge symmetry (unbroken at $T \sim M_{\Sigma} >> v_{EW}$) dictates that all the three triplet components (charged, doubly charged, and  neutral) contribute on equal footing, and are thus taken into account when computing the decay of $\Sigma$ into LL and the CP conjugate process.
 The choice of having more than one Higgs triplet is
essential to generate the lepton asymmetry \cite{HMS}. In this case,  the CP asymmetry in the decay of the lightest
Higgs triplet (which we choose to be $\Sigma_1$) is generated at one loop level due to the interference
between the tree and the one loop self energy diagram  and it is given by
\begin{equation}
\epsilon_{CP} \approx -\frac{1}{8\pi^2} \frac{\mbox{Im}\left[\mu_1\mu_2^*
\mbox{Tr}\left(\lambda^1\lambda^{2\dagger}\right)\right]}{M_2^2}
\frac{M_1}{\Gamma_1},
\end{equation}
where $\Gamma_1$ is the decay rate of the lightest Higgs triplet and it is given by
\begin{eqnarray}
\Gamma_1 = \frac{M_1}{8\pi}\left[\mbox{Tr}\left(\lambda^{1} \lambda^{1\dagger}\right)
+ \frac{\mu_1^2}{M_1^2}   \right].
\end{eqnarray}

We can compute the relevant traces in the DNTB$_0$ pattern to find:
\bea \label{traces} \mbox{Tr}(\lam^1 \lam^{2\dagger}) &=&
 2 (A_1-\frac{49}{47} B_1)(A^*_2-\frac{49}{47} B^*_2)+ (A_1+\frac{51}{47} B_1)(A^*_2+\frac{51}{47} B^*_2) \\
 \mbox{Tr}(\lam^1 \lam^{1\dagger}) &=& 2 |A_1-\frac{49}{47} B_1|^2 + |A_1+\frac{51}{47} B_1|^2
 \eea

If we denote now the phases of $A_a-\frac{49}{47}B_a,\;A_a+\frac{51}{47}B_a,\;\mu_a$ by $\a_a,\;\b_a,\;\phi_a$ ($a=1,2$) respectively, then by
rewriting the $A_a$'s and $B_a$'s coefficients of the $\lam^a$'s in the Yukawa term
($\lambda_{\alpha\beta}^{a}\, L_\alpha^T\, C\, \Sigma_a\, i\,\tau_2\, L_\beta$) in function of the combinations ($A_a-\frac{49}{47}B_a$)'s and
($A_a+\frac{51}{47}B_a$)'s, we find, say, the first combinations ($A_a-\frac{49}{47}B_a$)'s multiplied always with the field $\Sigma_a$. This means
that shifting the latter fields by a phase ($-\a_a$) would put the phases  $\a_a$ equal to zero.  For  $\mu_a \approx M_{\Sigma_a} \sim 10^{13}\, \mbox{GeV},a=1,2$ (which give a neutrino masses in the sub-eV range) we get:
\begin{equation}
\epsilon_{CP} \approx -\frac{1}{\pi} \frac{2|A_1-\frac{49}{47}B_1||A_2-\frac{49}{47}B_2|\sin(\phi_1 -\phi_2)+|A_1+\frac{51}{47}B_1||A_2+\frac{51}{47}B_2|\sin(\phi_1 -\phi_2+\b_1-\b_2)}
{1+2|A_1-\frac{49}{47}B_1|^2+|A_1+\frac{51}{47}B_1|^2},
\end{equation}

The baryon to photon density is approximately given by
\begin{equation}
\eta_B \equiv \frac{n_B}{s} =\frac{1}{3}\eta_L \simeq \frac{1}{3} \frac{1}{g_*}\,\kappa\, \epsilon_{CP},
\end{equation}
where $g_* \sim 100$ is the number of relativistic degrees of freedom at the time when the Higgs
triplet decouples from the thermal bath and $\kappa$ is the efficiency factor  which takes into account the fraction of out-of equilibrium decays and the washout effect. In the case of strong wash out , the efficiency factor can be approximated by ($H$ is the Hubble parameter)
\begin{eqnarray}
\kappa \simeq \frac{H}{\Gamma_1}(T= M_1),
\end{eqnarray}
 With the above numerical values and with an efficiency
 factor of order $10^{-4}$  we get, for $\b_1=\b_2$, a baryon asymmetry:
\begin{equation}
\eta_B\approx  10^{-7}\frac{
\mbox{Tr}\left(\lambda^1\lambda^{2\dagger}\right)}{\mbox{Tr}\left(\lambda^1 \lambda^{1\dagger}  \right) + 1}
\sin(\phi_2 -\phi_1).
\end{equation}
Thus one can bring about the correct baryon-to -photon ratio of $\eta_B \simeq 10^{-10}$ by choosing
$\lambda$'s of order $0.1$ and not too small relative phase between  $\mu_1$ and $\mu_2$.



\section{Summary and conclusion}
We have derived an explicit realization of the $U(1)$-symmetry underlying the non-tribimaximal pattern of the neutrino
mass matrix in the degenerate mass spectrum case. We deduced a bipartite form which characterizes uniquely this pattern. The departure from
the tribimaximal pattern is suggested by recent oscillation data, whereas the degenerate mass spectrum case is a good approximation
motivated by experimental data and numerical studies. One can consider it as a first step to be perturbed by a term proportional to the mass splitting ($m_2-m_1$), leading to a modified tripartite model without degeneracy. We have implemented this symmetry in a setup including
charged leptons supplemented either with many Higgs doublets or with many SM-singlet scalars. In both cases, one could accommodate the observed
charged lepton mass hierarchies. Similarly, the $U(1)$-symmetry can generate all sorts of neutrino mass hierarchies. We showed this explicitly
in type-I seesaw scenarios, where we found that no lepton/baryons asymmetry can be generated. However, in type-II seesaw mechanisms, one can account for the  photon/baryon density observed in the universe.

\section*{Acknowledgements}
 Part of the work was done
within the associate scheme program of ICTP.
N.C. acknowledges funding provided by the Alexander von Humboldt Foundation.
The work of S. N was  supported by the Algerian Ministry
of Higher Education and Scientific Research under the
PNR " Particle Physics/Cosmology: the interface", and
the CNEPRU Project No. D01720090023.



\bibliographystyle{mdpi}

\begin{thebibliography}{1}

\bibitem{nakamura10} K. Nakamura {\it et al.} (Particle Data Group), J. Phys. G {\bf 37}, 075021 (2010).

\bibitem{seesaw} M. Gell-Mann, P. Ramond and R. Slansky, in Supergravity, p. 315, edited by F. van Nieuwenhuizen
and D. Freedman, North Holland, Amsterdam, 1979;\\
T. Yanagida, Proc. of the Workshop on Unified Theory and the Baryon Number of the Universe,
KEK, Japan, 1979; \\
R. Mohapatra and G. Senjanovic, Phys. Rev. Lett. {\bf 44}, 912–915 (1980);\\
M. Magg and C. Wetterich, Phys. Lett. B
{\bf 94} (1980) 61.

\bibitem{morisi} S. Morisi and J. W. F. Valle, arXiv:1206.667,  to appear in Fortschritte der Physik - Progress of Physics;\\
S.-F. Ge, D. A. Dicus and W. W. Repko,
Phys.\ Rev.\ Lett. {\bf 108} (2012), 041801.


\bibitem{Xing} Z.Z. Xing; Phys. Lett. B {\bf 530} (2002), 159-166.



\bibitem{oztex} E. Lashin and N. Chamoun, Phys. Rev. D {\bf 85} (2012) 113011.

\bibitem{ntb} E. Lashin, M. Abbas , N. Chamoun and S. Nasri, Phys. Rev. D {\bf 86} (2012) 033013.


\bibitem{fogli} G. L. Fogli, E. Lisi, A. Marrone, A. Plaazzo and A. M. Rotunno,  Phys. Rev. D{\bf 84}, 053007 (2011).

\bibitem{HPS} P.F. Harrison, D.H. Perkins, and W.G. Scott,
Phys. Lett. B {\bf 530} (2002) 167; P.~F.~Harrison and W.~G.~Scott,
  Phys.\ Lett.\  B {\bf 535}, 163 (2002), Z.~z.~Xing,
  Phys.\ Lett.\  B {\bf 533}, 85 (2002).

\bibitem{matripar} E. Ma, Phys.\ Lett.\  B  {\bf 583} (2004) 157.


\bibitem{u1} E. Lashin, E. Malkawi , S. Nasri and N. Chamoun, Phys. Rev. D {\bf 83} (2011) 013002.



\bibitem{maform} E. Ma, Phys. Rev. Lett. {\bf 90}, 221802 (2003).

\bibitem{z2} E. Lashin, E. Malkawi , S. Nasri and N. Chamoun, Phys. Rev. D {\bf 80} (2009) 115013.


\bibitem{TB}
P.H. Frampton, S.T. Petcov and W. Rodejohann, Nucl.Phys. B {\bf 687} (2004) 31-54;
Guido Altarelli, Ferruccio Feruglio, Yin Lin, Nucl.Phys. B {\bf 775} (2007) 31-44;
I. de Medeiros Varzielas and G. G. Ross, Nucl. Phys. B {\bf 733} (2006) 31;
C. Luhn, S. Nasri and P. Ramond, Phys. Lett. B {\bf 652} (2007) 27;
R. N. Mohapatra, S. Nasri and H. B. Yu, Phys. Lett. B {\bf 639} (2006) 318.


\bibitem{Double} Y.~Abe {\it et al.}  [DOUBLE-CHOOZ Collaboration],
  Phys.\ Rev.\ Lett.\  {\bf 108}, 131801 (2012)
  [arXiv:1112.6353 [hep-ex]].

\bibitem{DAYA} F.~P.~An {\it et al.}  [DAYA-BAY Collaboration],
  Phys.\ Rev.\ Lett.\  {\bf 108}, 171803 (2012)
  [arXiv:1203.1669 [hep-ex]].

\bibitem{RENO}  J.~K.~Ahn {\it et al.}  [RENO Collaboration],
  Phys.\ Rev.\ Lett.\  {\bf 108}, 191802 (2012)
  [arXiv:1204.0626 [hep-ex]].

  \bibitem{Valle} D.~V.~Forero, M.~Tortola and J.~W.~F.~Valle,
  arXiv:1205.4018 [hep-ph].




\bibitem{SV} J.~Schechter and J.~W.~F.~Valle,
  Phys.\ Rev.\ D {\bf 25}, 774 (1982).

\bibitem{W} L.~Wolfenstein,
  Phys.\ Rev.\ D {\bf 26}, 2507 (1982).

\bibitem{ma03} E. Ma and G. Rajasekaran, Phys. Rev. D {\bf 68} (2003) 071302(R).

\bibitem{BjorWein77} J. D. Bjorken and S. Weinberg, Phys. Rev. Lett. {\bf 38}, 622 (1977).

\bibitem{FY} M. Fukugita and T. Yanagida, Phys. Lett. B {\bf 174} (1986) 45.

\bibitem{KRS} V.~A.~Kuzmin, V.~A.~Rubakov and M.~E.~Shaposhnikov,
  Phys.\ Lett.\  B {\bf 155}, 36 (1985).

\bibitem{HMS} E.~Ma and U.~Sarkar,
  Phys.\ Rev.\ Lett.\  {\bf 80}, 5716 (1998); T.~Hambye, E.~Ma and U.~Sarkar,
  Nucl.\ Phys.\  B {\bf 602}, 23 (2001).




\end{thebibliography}

\end{document}